\documentclass[sigconf,authorversion,nonacm]{acmart}

\usepackage{fancyhdr}
\AtBeginDocument{%
    \addtolength{\footskip}{2.0\baselineskip}%
    \fancyfoot[L]{\textit{\textbf{Preprint --- final version forthcoming}}}%
}

\usepackage[utf8]{inputenc} %
\usepackage[T1]{fontenc}    %
\usepackage{hyperref}       %
\usepackage{url}            %
\usepackage{booktabs}       %
\usepackage{amsfonts}       %
\usepackage{nicefrac}       %
\usepackage{microtype}      %
\usepackage{xcolor}         %
      \let\cite\citep
      
\usepackage{tabularx}
\usepackage[noend]{algorithm2e}
\LinesNumbered
\RestyleAlgo{ruled}
\usepackage{enumitem}
\usepackage{pgf,tikz}
\usetikzlibrary{positioning}
\usepackage{tikz-network}
\usepackage{dsfont}
\usepackage{subcaption}
\usepackage{amsmath}
\usepackage{cleveref}
\crefname{algocf}{alg.}{algs.}
\Crefname{algocf}{Algorithm}{Algorithms}
\usepackage{booktabs}
\usepackage{amsthm}
\usepackage{textcomp}

\AtBeginDocument{%
  \providecommand\BibTeX{{%
    \normalfont B\kern-0.5em{\scshape i\kern-0.25em b}\kern-0.8em\TeX}}}

\LinesNumbered
\RestyleAlgo{ruled}
\crefname{algocf}{alg.}{algs.}
\Crefname{algocf}{Algorithm}{Algorithms}

\begin{document}

\title{Algorithmic Collective Action with Two Collectives}

\author{Aditya Karan}
\email{karan2@illinois.edu}
\affiliation{%
  \institution{University of Illinois at Urbana Champaign}
  \country{USA}
}

\author{Nicholas Vincent}
\email{nvincent@sfu.ca}
\affiliation{%
  \institution{Simon Fraser University}
  \country{Canada}
}

\author{Karrie Karahalios}
\email{kkarahal@illinois.edu}
\affiliation{%
 \institution{University of Illinois at Urbana Champaign}
  \country{USA}
}

\author{Hari Sundaram}
\email{hs1@illinois.edu}
\affiliation{%
 \institution{University of Illinois at Urbana Champaign}
  \country{USA}
}

\renewcommand{\shortauthors}{Karan et al.}

\begin{abstract}

Given that data-dependent algorithmic systems have become impactful in more domains of life -- from finding media to influencing hiring -- the need for individuals to promote their own interests and hold algorithms accountable has grown. The large amount of data these systems use means that individuals cannot impact system behavior by acting alone. To have meaningful influence, individuals must band together to engage in collective action. The groups that engage in such \textit{algorithmic collective action} are likely to vary in size, membership characteristics, ability to act on data, and crucially, objectives. In this work, we introduce a first of a kind framework for studying collective action with two or more collectives that strategically behave to manipulate data-driven systems. With more than one collective acting on a system, unexpected interactions may occur. 
We use this framework to conduct experiments with language model-based classifiers and recommender systems where two collectives each attempt to achieve their own individual objectives. We examine how differing objectives, strategies, sizes, and homogeneity can impact a collective's efficacy. We find that the unintentional interactions between collectives can be quite significant. We find cases in which a collective acting in isolation can achieve their objective (e.g., improve classification outcomes for themselves or promote a particular item), but when a second collective acts simultaneously, the efficacy of the first group drops by as much as 75\%.
We find that, in the recommender system context, neither fully heterogeneous nor fully homogeneous collectives stand out as most efficacious and that the impact of heterogeneity is secondary compared to collective size. 
Our results signal the need for more transparency in both the underlying algorithmic models and the different behaviors individuals or collectives may take on these systems. This approach also allows collectives to hold algorithmic system developers accountable and illustrates a framework for people to actively use their own data to promote their own interests.

\end{abstract}


\maketitle

\section{Introduction}

Collective action campaigns in which people change their data generating behaviors can impact operations of algorithmic systems \cite{vincent_data_2021,wu_reasonable_2022}. 
As an example, consider several very differently motivated types of collective action on the same platform. Perhaps there is a fan group of a small artist that wants to manipulate a recommender system to promote their work \cite{Nugent_2021}. Another group might coalesce with the goal of demoting content from a controversial figure \cite{payne2024review}. Yet another group might be interested in broadly harming the platform by reducing recommender system performance on the basis of some platform policy (\textit{e.g.} policies on AI art \cite{jiang_ai_2023, shan2024nightshade, devon_billie_2024}). All of these groups might independently act by manipulating their data flow (interacting with or rating items from a certain artist, increasing or decreasing platform usage, etc.), unaware of each other's actions, and resulting in cases where their actions clash or synergize. 

Public opinion polls suggest growing appetite for collective action \cite{robertson_exclusive_2024,the_authors_guilds_new_2023}, and it is only inevitable that \textit{multiple} campaigns, each with their own goals, will emerge that target the same algorithmic system. These collectives may be unaware of each other, and the complexity of modern ML systems makes it possible that campaigns that have seemingly distinct goals can inadvertently interact by changing the weights of a model during training.

Given that the motivation to engage in such action is likely to further increase, there is an urgent need to understand the dynamics of such collective action, from the perspective of organizers trying to be efficacious in allocating resources toward social movements, AI developers trying to make product decisions accordingly, and the public who are becoming increasingly aware of how algorithms may be manipulated.

Prior work has looked at how a small group can effectively influence algorithmic systems, including early recommender systems \cite{Dellarocas2000-tt,Lam2004-me} and modern deep learning systems  \cite{Guo2023-jt}. This research has considered ``shilling'' (posting fake positive reviews to promote a service) a set of items in a ranking system \cite{si2020shilling} and a malicious attacker who wishes to manipulate a system \cite{tian2022comprehensive, lin_ml_2021}. However, no work has tackled cases where more than one collective, each with their own objectives, engages with the same algorithmic system.

In classical collective action,~\citet{ostrom1990governing} notes that small, homogeneous groups of people with ability to sanction are able to effectively overcome some hurdles of collective action. In algorithmic settings, however, it is not clear what types of groups (small vs large or homogeneous vs heterogeneous) most effectively achieve some objective.
In this work, we develop a framework for understanding simultaneous collective action campaigns by distinct groups. Different collectives likely will have different goals and may engage in structurally different actions. We first consider how these different objectives, strategies, and sizes impact the ability for each collective to achieve their objective (\textbf{RQ1}). We then consider how different collectives might form and how collective homogeneity impacts their efficiency (\textbf{RQ2}). 

Using this framework, we conduct experiments with language model-based classifiers and recommender systems. We find that these interactions can be quite significant, \textit{e.g.}, two collectives that were each able to successfully change a language model's prediction to their favor with near 100\% efficacy when acting alone can drop to nearly 25\% efficacy when other collectives act simultaneously (\textbf{RQ1}). We find homogeneity as a secondary influence compared to group size in the recommender system task (\textbf{RQ2}). These examples highlight the need to study collective action with multiple distinct collectives going forward. Our contributions are as follows:

\begin{description}[style=unboxed,leftmargin= .35cm]
\item[Establishing a Framework for Algorithmic Collective Action by Multiple Collectives:]  We create a first of a kind framework for analyzing multiple collectives acting on an algorithmic system. Prior work \cite{hardt_algorithmic_2023, baumann2024algorithmic} has focused on a single collective scenario. The key insight underlying our framework is that each collective is likely to have their own objectives, targets, and capabilities. Our framework helps analyze any type of system where user provided data is used. We examine the role of differing objectives, strategies, collective size, and homogeneity while describing other aspects for future research.

\item[Insights from Experiments with Multiple Collectives Taking Action with Data:] Using our framework, we establish initial findings about between-group interactions. While prior works have studied single collectives with a single objective, here we consider when multiple collectives each have their own objective. We find certain strategies can lead to unexpected, unintentional interactions between collectives that simultaneously undermine their objectives. We find that collective size plays a significant role in efficacy, and that homogeneity can play a secondary influence. The results show the underlying complexity of interactions among distinct collectives with differing objectives and demonstrate the need to further understand how simultaneous strategic behavior can impact algorithmic systems.
\end{description}

\section{Related Work}

\textbf{Collective Action with Algorithms and Data:} ~\citet{hardt_algorithmic_2023} defines the notion of algorithmic collective action in a stylized model, assuming one group and examining the group size that is needed to effect changes from a theoretical perspective as well as empirically, using language models to classify text. ~\citet{baumann2024algorithmic} examine this framing in the song recommendation context. These works consider just a single collective and assume that the impact of the collective is related just to the size of the group, rather than the users themselves. In many applications, however, the user's own prior history and behavior with the algorithmic system can greatly influence the impact that this user has. In classical collective action, ~\citet{ostrom1990governing} notes certain criteria (size and homogeneity) as key factors in overcoming challenges of collective action, however, it is not clear whether this holds in algorithmic settings. Moreover, the presence of multiple collectives with varying attributes relevant to algorithmic systems is a new perspective we seek to address. 

\textbf{Understanding and Countering Impacts of Algorithms:} 
A number of scholars in computing have sought to understand how people engage in forms of algorithmic resistance, using lenses such as folk theory \cite{Karizat2021-we,DeVito2017-jv,devito2018algorithm} and auditing \cite{Shen2021-pq,devos2022toward}. This work is heavily predicated on the need to understand how users reason about the algorithmic systems they interact with. 
Others have looked at domain specific contexts such as online content marketplaces ~\cite{etter_activists_2021, bandy_tulsaflop_2020}, ridesharing ~\cite{lei_delivering_2021, wells_just--place_2021, woodside_bottom-up_2021, jeroen_platform_2021, hastie_platform_2020} and forms of activism ~\cite{shaw_computer_2014, newlands_collective_2018}. Some work has looked at manipulation and data campaigns to promote pro-social outcomes ~\cite{malchik_data_2022, vincent_can_2021, vincent_data_2021,wu2022reasonable}
while ~\cite{hardt_strategic_2016, milli_social_2019} have looked at the impact of strategic manipulation. However, none of them have looked at multiple collectives with different strategic interests. Our work extends these domains by looking at what happens with multiple different interests operating on algorithmic systems. The multiple collectives framing provides an additional dimension for both users and researchers to consider.

\textbf{Shilling and Adversarial Machine Learning:} From a technical perspective, our approach looks at similar scenarios to shilling \cite{si2020shilling} (where fake positive reviews are used to boost a product, often in exchange for money) and adversarial ML. Most adversarial literature assumes that attacks are malicious, but some work ~\cite{feng2022has} uses adversarial methods to improve fairness outcomes. 
In an adversarial ML lens, our work is most closely situated in the targeted poisoning, where the goal is to alter training data to get a specific outcome, potentially on a specific part of the population ~\cite{tian2022comprehensive, lin_ml_2021}. 
More recently, subpopulation data poisoning, where only a subset of the data is altered to influence outcomes on a subpopulation, has been explored ~\cite{jagielski2021subpopulation, gupta2024fragilegiantsunderstandingsusceptibility}. However, most work along these lines focus on how to achieve a specific target by crafting the optimal attack technique. Our work focuses on \textit{who} is trying to attack the ML system and how this may determine their objectives and efficacy. In addition, we also examine how \textit{multiple} groups acting on a system, each with their own group compositions, can impact models. While some adversarial work has looked at multi-objective scenarios \cite{deng_multi-objective_2019, williams_black-box_2023, garcia_learning_2020}, to our knowledge, no work specifically tackles multiple groups with differing objectives acting upon an algorithmic system.

\section{Framework}
\begin{figure*}
    \includegraphics[width=\linewidth]{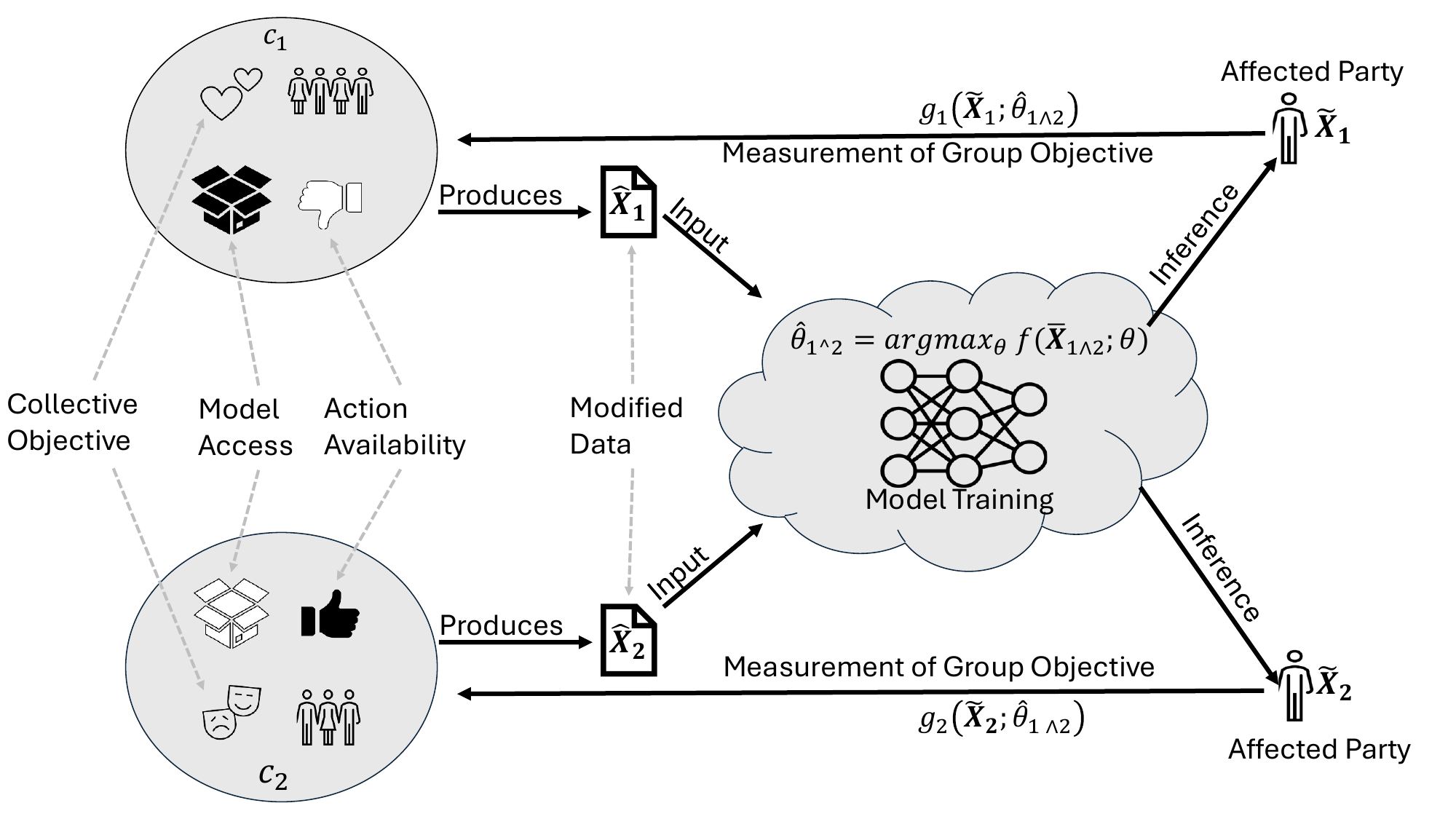}
    \caption{Framework overview. Collectives $c_1$ and $c_2$ form with different objectives. The items they aim to modify (romance vs drama), access to the model (white-box vs black-box access), actions they can do (promote vs demote) may vary. With this, the collectives produces $\hat{X_i}$, their individually modified dataset. Both collectives send their individually modified data into the model, which is then combined to produce $\bar{\mathbf{X}}_{1 \wedge 2}$. The system's objective function produces parameters $\hat{\theta}_{1 \wedge 2}$ which is influenced by the combined actions of $c_1$ and $c_2$. The collectives measure their success by their objective functions ($g_i$) applied on $\tilde{X_i}$, their target dataset, with the common parameters $\hat{\theta}_{1 \wedge 2}$}
    \label{fig:multiple_collective_framework}
\end{figure*}

In expanding from a single collective to multiple, we must critically examine \textit{how} collectives may differ from each other. This allows us to specifically articulate distinguishing features of any single collective, and how these distinct collectives may end up interacting. 
Here we establish a framework to identify the components of algorithmic collective action with multiple collectives. \Cref{fig:multiple_collective_framework} summarizes the key components of this framework: Number of Collectives/Objectives, Collective Construction, Action Availability, Model Access, Affected Parties, and Measurement. 

\subsection{Framework Components}

\begin{description}[style=unboxed,leftmargin= .35cm]
\item[Number of Collectives and Objectives:]
Prior work on collective action  \cite{hardt_algorithmic_2023,vincent_data_2021, baumann2024algorithmic}, adversarial machine learning, and shilling has broadly focused on the case where just one collective tries to achieve a specific goal. However, as more collectives seek to ensure that the outcome of these algorithms suit their own purposes, different collectives, each with their own objective, will start to emerge. 

\item[Collective Construction/Properties:] 
In practice, different users have different abilities to influence a model. This can come because the data that one can control is different for each user (\textit{e.g.}, users can only modify their own prior ratings). The collectives can also form from different processes(\textit{e.g.,} formed from fans of a particular artist vs at a workplace). The sizes of these collectives can vary and may be able to affect some parts of a model more than others. Prior work samples users uniformly \cite{hardt_algorithmic_2023, baumann2024algorithmic}.

\item[Action Availability:] Different algorithmic systems have different types of actions available to them (ex. rating a movie, posting positive sentiment comments). Users may be able to add entries or alter existing ones to affect the model. In other scenarios, users may be able to affect other entries that they are typically not allowed to. 

 \item[Access to Model:] A collective's access to the model can inform the types of actions they can perform. If the collective has white-box access or the model is public, a collective can construct a highly specific adversarial attack, while if a collective has no access and no awareness of models, they may have to rely on more unsophisticated tactics. 
 
\item[Affected Party:]  A collective can choose to influence outcomes for the entire dataset or a subset. For example, a fan group of a specific artist may try targeting users who are interested in the same genera or the broader community. \cite{hardt_algorithmic_2023} affects users with a specific modification in a resume while \cite{baumann2024algorithmic} affects all users. In data activism cases, where the goal is to influence certain policies, targeting the action to narrowly target a specific set of users can be more effective than trying to change the outcomes more broadly.

\item[Measurements:] A collective's ability to measure their efficacy will impact whether the collective can continue or adjust tactics. Various different fields such as psychology ~\cite{overington_progress_2012}, education ~\cite{albon_mutual_2014}, political science ~\cite{shadmehr_collective_2011} and management ~\cite{burns_using_2008, rosen_team_2017, widmeyer_team_1997} all consider the role of measurement, uncertainty and mutual progress tracking. In this context, there is a challenge on whether the collectives can measure their outcomes themselves, and if so, how does feedback on their progress influence their subsequent actions. 

\end{description}

As an example, we may have two collectives, one wants to promote artist A while another wants to demote artist B. The first collective forms from a fan community, while the second forms from people who are opposed to certain statements artist B made. The first collective may make positive rating changes to all artists associated with A, while the second makes negative ratings only to artist B. The first collective may have no knowledge of the underlying system, while some members in the second collective have some access to the underlying model. The first collective wants to promote artist A to people who are fans of similar related work, while the second may want to demote this artist wholesale. The first collective measures success based on increasing unique users consumed content from artist A while the second collective measures success by decreasing the number of users consuming content from B.

\subsection{Collective Archetypes}
Different collectives can have different objectives, targets, and means of measurement. To simplify our discussion, we will define two primary archetypes that we will explore in our experiments. These archetypes are defined in relation to their intended goal or what downstream behavior they want to cause the system to do. 

\begin{description}[style=unboxed,leftmargin= .35cm]
\item[Targeted Promoter:] This collective's focus is to promote a specific item or set of items, causing them to be treated more favorably by a machine learning system. This could involve causing a specific item to be ranked higher in a recommender system, causing members of a collective to be classified in a certain manner among other outcomes, or causing certain tokens to show up more frequently in a language model's outputs.

\item[Targeted Demoter:] This collective's focus is to demote a specific object so that it's treated unfavorably by the machine learning system. This could involve causing a specific movie to be ranked lower in a recommender system or causing members of a collective to be classified unfavorably. 

\end{description}

Other archetypes that are important considerations are a ``broad based promoter'' (a collective that broadly wants to improve the ML system) or a ``broad based demoter'' -- a collective that broadly wants an ML system to perform poorly across the board. We choose in this work to focus on the former two archetypes, as these most directly represent the targeted, competing interest that may not directly conflict with the ML system's objective. This type of behavior has been observed in media recommendation \cite{Nugent_2021} and online reviews \cite{anderson2012learning, luca2016reviews}. We note that these archetypes are ``soft''. An actual collective's objective can be more complex than these archetypes. Nevertheless, they allow us to more readily categorize possible sets of a collective's motivations.

\section{Experiments}

We use our framework to conduct several experiments to investigate how one collective may impact the objective of another collective, whileexamining how different collective archetypes may impact each other. We examine two research questions (RQs), focusing first on interaction effects that can come from different overlaps in strategy, targets and size of collectives in language model based classification (\Cref{subsec:classifcation_experiment} and \Cref{subsec:classification_results}). We then investigate the role of collective heterogeneity using the context of recommender systems (\Cref{sec:recsys_experiment} and \Cref{subsec:recsys_results}). We further define collective heterogeneity in the experimental setup: at a high level, this refers to the how ``similar'' members of the collective are to each other. 
We use a recommender for the second RQ as recommender data is more well-suited for studying complex collective formation. Our research questions are as follows.

\begin{description}
    \item [RQ1:] When two collectives have their own objectives and are acting on a system, how do different targets, strategies and collective sizes affect the ability of each collective to achieve their respective objective?
    \item [RQ2: ]
    When two collectives each have their own objectives, how does collective heterogeneity affect their ability to achieve said objective?
\end{description}

To futher help build intuition, we also examine a linear case in \Cref{sec:appendix_linear}.

\subsection{Language Model Experimental Setup}
\label{subsec:classifcation_experiment}

We expand \citet{hardt_algorithmic_2023} to a two collective setting. Here, a finetuned BERT model performs multiclass prediction on resumes ~\cite{jiechieu2021skills} to predict the roles of the candidates. The collective members plant a ``signal'' in their input resumes as well as adjust their classification labels (as done in prior work). This modified data is then used as part of training. The collective's goal is to ensure that when said signal appears at evaluation time, the model will classify resumes with the signal to the desired class. Here, we consider cases where multiple ``targeted promoters'' collectives modify their input texts (resumes) and their training labels to influence the downstream classification. The collective is successful if the top-one prediction accuracy of resumes sharing the same signal is high -- this is the same as one of the metrics used in \citet{hardt_algorithmic_2023} (full experimental setup details in \Cref{sec:appendix_classification}). 

For our experiments, we consider a two collective scenario where each collective plants a signal. Each collective's goal is that all data points with the same planted signal will be classified to its desired target class $t_i$. Each collective may employ their own signal to achieve their desired result. In our framework, we can express the objective $g_i$ as the probability of predicting target class $t_i$ on target data 
$\mathbf{\tilde{X}}_i$ given language model's learned parameters $\theta$ - \textit{i.e.,} $g_i(\mathbf{\tilde{X}}_i |  \theta)= p(\mathbf{\tilde{X}_i} = t_i ; \theta)$. We can compare the outcome when two collectives interact on the system, $g_i(\mathbf{\tilde{\mathbf{X}}}_i |  \theta_{i \wedge j})$ with learned parameters  $\theta_{i \wedge j}$, vs a single collective, $g_i(\mathbf{\tilde{X}}_i |  \theta_{i})$, with learned parameters $\theta_{i}$. Here, the collective's only care about efficacy on resume's planted with the specific signal (modification).

We note that this type of strategy/targeting is a more specific instance of the general actions available in a collective action setting. While we choose to focus on a very specific type of strategy (modification of text resume data), this approach can be readily applied to any number of models, targets and behaviors.

\subsection{Recommender Systems Experimental Setup}
\label{sec:recsys_experiment}
\textbf{Scenario and Assumptions:}
For this set of experiments, we model two collectives that can be either a ``targeted promoter'' or a ``targeted demoter'' archetype - \textit{i.e.} they either want to promote or demote specific items. These users do not have direct model access, and the only action they can do is either add or change their rating on a movie they want to directly impact. We focus on this case, as these types of behavior (promoting or demoting specific objects), has been observed in various contexts ~\cite{etter_activists_2021, bandy_tulsaflop_2020, si2020shilling}. We assume that the ``target items'' of any given campaign are a function of the collective members. In our experiments, the actions available to participants are either changing a rating to the highest or lowest possible rating. We assume that the collective wants to change the overall ranking of a set of items across all users. We form collectives by first grouping users into initial clusters based on some notion of similarity, which we instantiate for our experiments in \Cref{subsec:recsys_results}. We generate our collective by sampling from these clusters. By controlling the propensity of sampling from a seed cluster, we can control the homogeneity of the resulting collective.

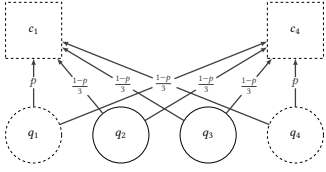
\begin{figure}
        \resizebox{0.5\columnwidth}{!}{%
            \begin{tikzpicture}[scale=.7]
                \tikzset{
                    cluster/.style={minimum size=2.25cm, draw=black, fill=white},
                    clusterchosen/.style={cluster, dashed},
                    edge/.style={
                        Direct,
                        NotInBG=True,
                        line width=2pt %
                    }
                }

                \Vertex[x=0, y=1.59, size=.1, style={color=white}, position=above, color=white]{pseudo}

                \Vertex[x=0, y=6, size=2.25, shape=rectangle, color=white, style=clusterchosen, label = {\huge $c_1$}]{c1}
                \Vertex[x=15, y=6, size=2.25, shape=rectangle, color=white, style=clusterchosen, label = {\huge $c_4$}]{c2}

                \Vertex[x=0, y=0, size=2.25, shape=circle, color=white, style=clusterchosen, label = {\huge $q_1$}]{q1}
                \Vertex[x=5, y=0, size=2.25, shape=circle, color=white, style=cluster, label = {\huge $q_2$}]{q2}
                \Vertex[x=10, y=0, size=2.25, shape=circle, color=white, style=cluster, label = {\huge $q_3$}]{q3}
                \Vertex[x=15, y=0, size=2.25, shape=circle, color=white, style=clusterchosen, label = {\huge $q_4$}]{q4}
                \Edge[Direct, NotInBG = True, label={\huge $p$}](q1)(c1)
                \Edge[Direct, NotInBG = True, label={\huge $\frac{1-p}{3}$}](q2)(c1)
                \Edge[Direct, NotInBG = True, label={\huge $\frac{1-p}{3}$}](q3)(c1)
                \Edge[Direct, NotInBG = True, label={\huge $\frac{1-p}{3}$}](q4)(c1)

                \Edge[Direct, NotInBG = True, label={\huge $\frac{1-p}{3}$}](q1)(c2)
                \Edge[Direct, NotInBG = True, label={\huge $\frac{1-p}{3}$}](q2)(c2)
                \Edge[Direct, NotInBG = True, label={\huge $\frac{1-p}{3}$}](q3)(c2)
                \Edge[Direct, NotInBG = True, label={\huge $p$}](q4)(c2)

            \end{tikzpicture}%
        }
        \captionsetup{width=\linewidth}
        \caption{Group construction process used for forming two collectives ($C=2$). Here we have $Q= 4$ clusters of users. $q_1$ serves as the seed for $c_1$, and $q_4$s serves as the seed for $c_4$. The collectives are then constructed by sampling with probability $p$ from their seed cluster and then uniformly at random from the remaining clusters $\frac{1-p}{3}$.}
        \label{fig:group_construction}
\end{figure}

\textbf{Collective Formation and Item Selection:}
To form collectives, we start with initial recommender system data $M$. We first use a matrix factorization model, which decomposes the data $M = U^TV$ where $U$ are the user vectors and $V$ are the item vectors. These user vectors can be used as features to cluster similar users together. We cluster users into $Q$ clusters based on their user vectors. This gives us a purely data driven approach to clustering users. We expect that users that have similar interests are more likely to interact with each other if embarking on a data campaign. This results in $Q$ clusters $(q_1, q_2, ... q_Q)$. We pick $C$ clusters to use as seeds for our eventual collective. These clusters can be picked uniformly at random or picked to be maximally distant from each other. We will call $c_i$ the collective that has seed cluster $q_i$.
To assign users to a collective, we introduce a sampling propensity $p$ which represents the probability of picking a user for collective $c_i$ from cluster $q_i$ (\Cref{fig:group_construction}). A propensity of $1$ means that $c_i$ would only have users from $q_i$ while a propensity of $\frac{1}{Q}$ means that $c_i$ will have roughly equal proportion of users from all clusters. By varying $p$ we can examine different levels of homogeneity from completely homogeneous ($p = 1$) to fully heterogeneous. Understanding the role of this homogeneity is crucial, as collectives in the real-world form from various processes, resulting in collectives having differing levels of homogeneity. 
After seed clusters are chosen, we sample uniformly at random from users in the chosen cluster. 
We repeat this process until we get $N$ users for every $c_i$ cluster. Once we have the collective members, we determine the item set that they will act on. We look at the collectively highest rated items and select $V$ items. If a collective is highly similar ($p$ is close to $1$) we expect top-rated items to also be highly similar.

\textbf{Evaluation:}
To evaluate effectiveness of the collective's behavior, we measure the relative hit ratio of the items targeted by the collectives for all users in the test set. We let $g_{i}$ (the objective function for collective $c_i$) be the Hit Ratio at k (HR@k or Hit Ratio) for a set of items chosen by collective $i$. The HR@k ratio measures what fraction of users see a given item or set of items in their top $k$ rankings, a higher HR@k ratio means more users are seeing the item of interest. ``Promoter'' collectives want the ratio to increase, while a ``demoter'' collectives will want the ratio to decrease.

Let the collective's target to be $\tilde{\mathbf{X}}_i$. Let $\theta'$ be the parameters learned from the model without any modifications and $\hat{\theta}_i$ be the parameters learned when $c_i$ performs its modification strategy. For this set of experiments, we assume the collective is trying to affect all potential users.
 We let $g_{i}(\mathbf{\tilde{X_i}}|\theta')$ be $c_i$'s objective without any collective modifying ratings, while $g_{i}(\mathbf{\tilde{X_i} | \hat{\theta_i}})$ evaluates $c_i$'s objective when performing their collective action. The relative hit ratio can be written as 
 $\frac{g_i(\tilde{X_i}|\hat{\theta_i})} {g_i(\tilde{X_i}|\theta')}$. 
Collectives that have a goal to increase the relative ranking want the HIT ratio to be greater than 1, while ones that want to decrease the relative ranking want the relative hit ratio to be less than $1$. 
To measure how different collectives affect each other, we can compute 
$g_{i}(\mathbf{\tilde{X}}_{i}  | \hat{\theta}_{i \wedge j})$  where $\hat{\theta}_{i \wedge j}$ is the learned when both collectives $i$ and $j$ collectives make their modifications. We define the constructiveness of collective $j$ on $i$'s objective as $CT(c_i, c_j) = \frac{g_{i}(\tilde{X_{i} }| \hat{\theta}_{i \wedge j})} {g_i(\tilde{X_{i}} | \theta') } -  \frac{g_{i}(\tilde{X_{i}} | \hat{\theta}_{i})} {g_i(\tilde{X_{i}} | \theta') }$. When the constructive score is positive, $c_j$ is constructively boosting $c_i$ and if the score is negative, $c_j$ is interfering with $c_i$.

\section{Results}

We first analyze the role of differing objectives, target, strategies, and group sizes in the two collective scenario in language model classification in \Cref{subsec:classification_results}. We then further examine the role of collective homogeneity in \Cref{subsec:recsys_results}.

\subsection{RQ1 Role of Multiple Objectives}
\label{subsec:classification_results}

\textbf{Experimental Details:} 
Following  ~\citet{hardt_algorithmic_2023}, we finetune \texttt{distilbert-based-uncased} ~\cite{Sanh2019DistilBERTAD} using Hugging Face ~\cite{wolf2020huggingfaces} for five epochs with default hyperparameters. Full details on the training setup can be found in \Cref{sec:appendix_classification}. Each collective in this case is a targeted promoter; they perform their modifications to achieve high top one accuracy prediction on their desired target class using a specific strategy. All strategies in these experiments are of the same form: inserting a specific character (denoted by the number) every 20 words. The difference lies in the specific character inserted. Collectives are differentiated by their target and strategy. We denote each of these collective archetypes as $[Letter][Number]$ where the letter refers to a target class and the number refers to a specific modification strategy (\textit{i.e.} which character to use). For example, if we compare collectives $A1$ and $A2$ we are referring to two collectives which have the same target class but employing a different strategy. The specifics instantiations of these collectives can be found in \Cref{sec:appendix_classification}. In our experiments, we consider two strategy pairings (``0" vs ``1" and ``100" vs ``101"). We examined two sets to understand whether high level semantic similarities/difference between these strategies may play a role. In particular, ``0" vs ``1" are two characters that are semantically similar, and appear several times in the dataset $(<30$ instances over $25,000$  data points) while ``100" vs ``101" are both characters never seen in the dataset. The mapping of these numbers to characters can be found in \Cref{sec:appendix_classification}.

\textbf{Evaluation:}
With our two strategy sets, we consider three scenarios for each of them. The first is when each collective targets different classes with separate strategies. The second is where both collectives use the same strategies but different target classes. The third is where both collectives use different strategies but the same target class. The objective $g_i$ is to have a high probability of predicting the target class. The set of resumes to evaluate on $\tilde{X}_i$, resumes with the prescribed modification. We directly plot $g_i(\mathbf{\tilde{X}_i} | \hat{\theta}_{i\wedge j})$ (collective i's objective when both i and j are acting on the system) as well as $g_i(\mathbf{\tilde{X}_i} | \hat{\theta}_{i})$ (collective i's objective when just i is acting on the system).

\begin{figure*}[t]
    \centering
    \begin{subfigure}{0.31\textwidth}
        \centering
        \includegraphics[width=\linewidth]{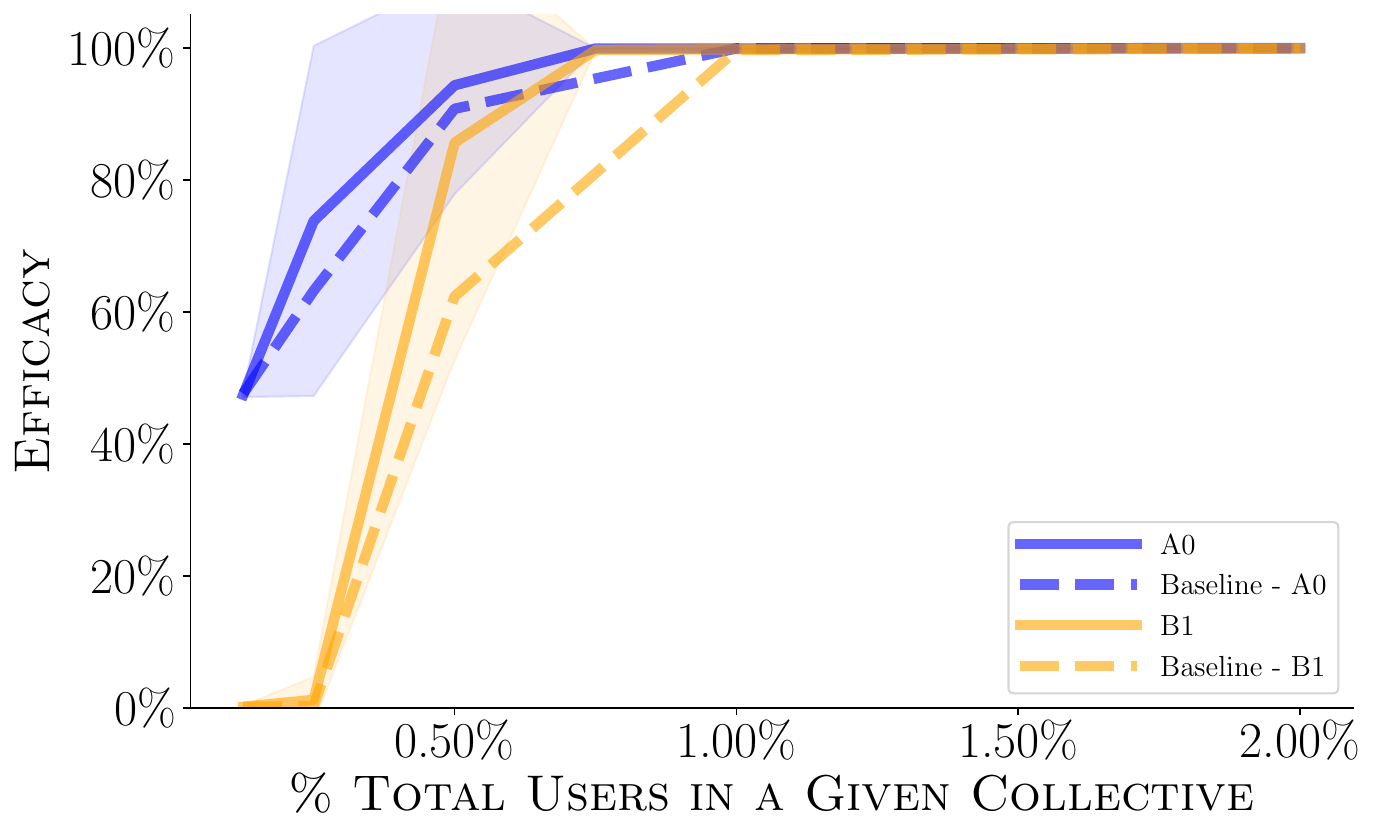}
        \caption{Different targets \& chars }
        \label{fig:normal_char_two_seperate}
    \end{subfigure}
    \begin{subfigure}{0.31\textwidth}
        \centering
        \includegraphics[width = \linewidth]{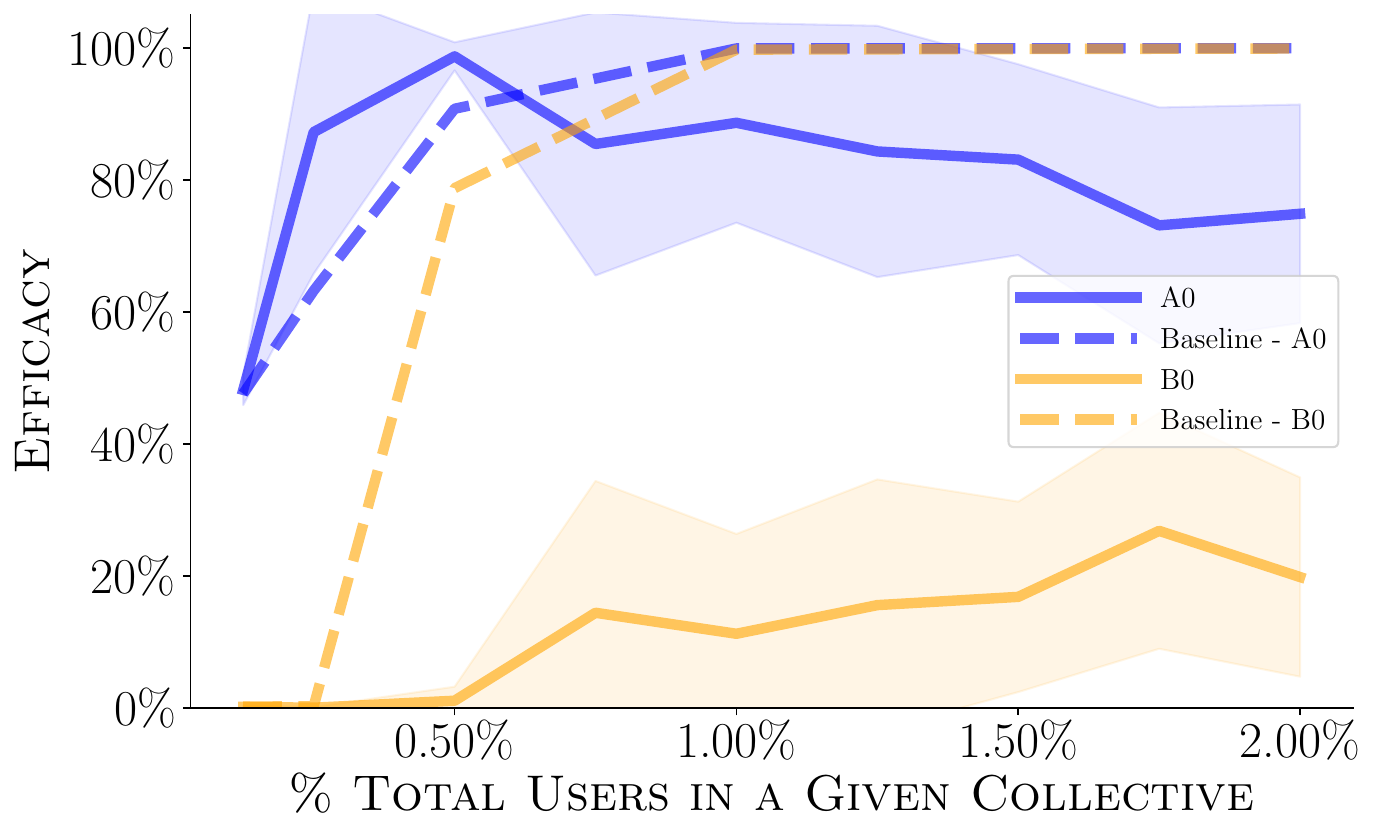}
    \caption{Different targets, Same chars}
        \label{fig:normal_char_same_char}
    \end{subfigure}
    \begin{subfigure}{0.31\textwidth}
        \centering
        \includegraphics[width=\linewidth]{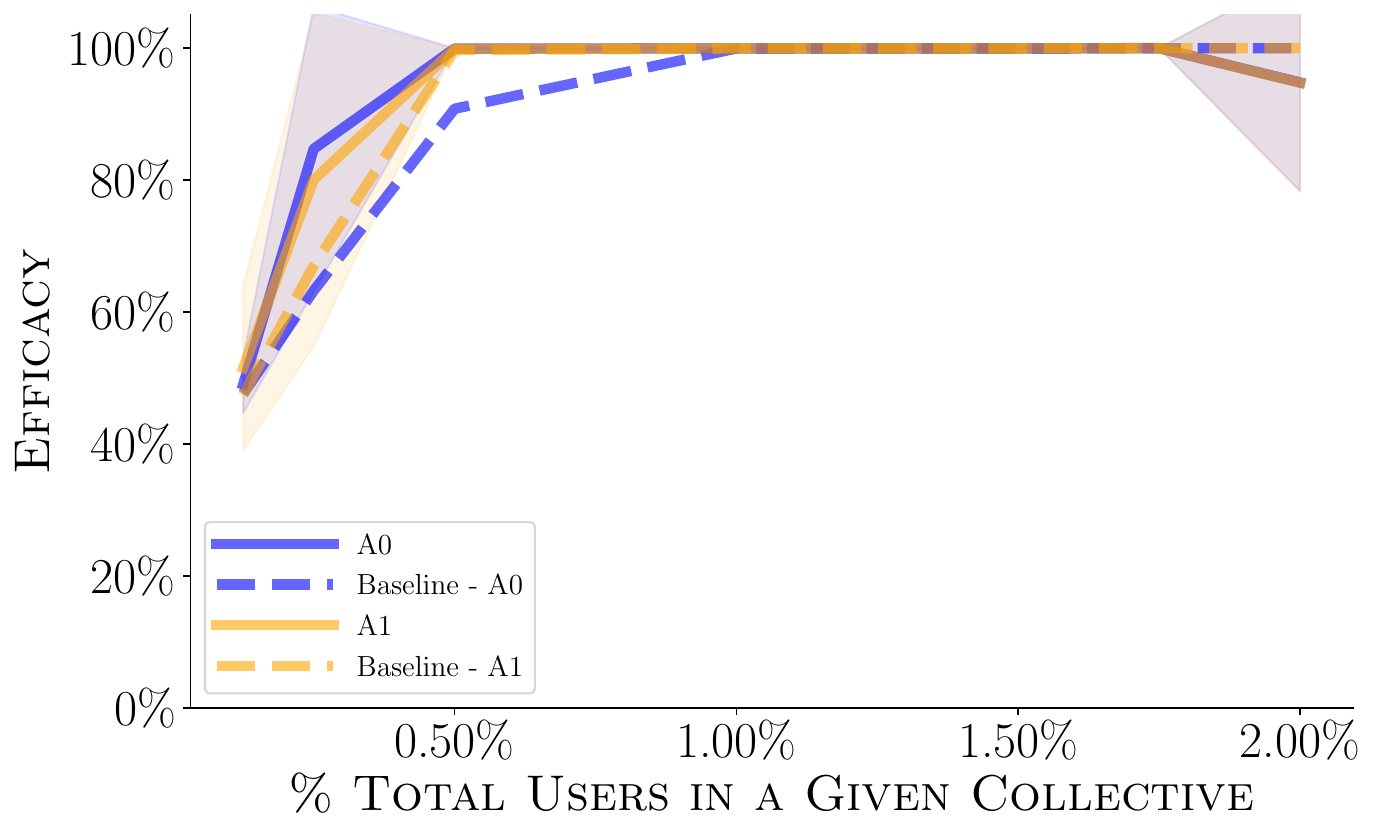}
        \caption{Same targets, Different chars}
        \label{fig:normal_char_same_target}
    \end{subfigure}

    \begin{subfigure}{0.31\textwidth}
        \centering
        \includegraphics[width=\linewidth]{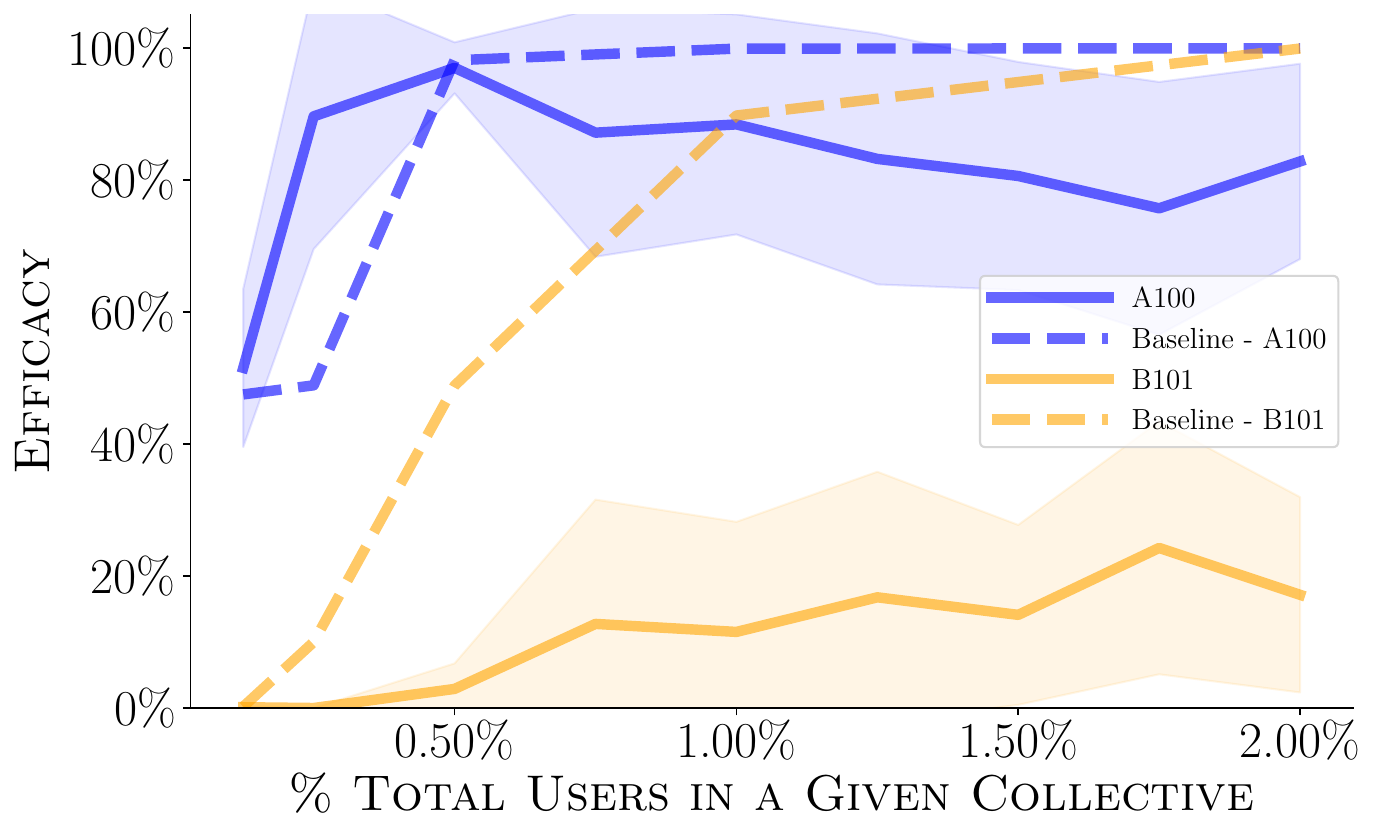}
        \caption{Different targets \& chars}
        \label{fig:special_char_two_seperate}
    \end{subfigure}
    \begin{subfigure}{0.31\textwidth}
        \centering
        \includegraphics[width = \linewidth]{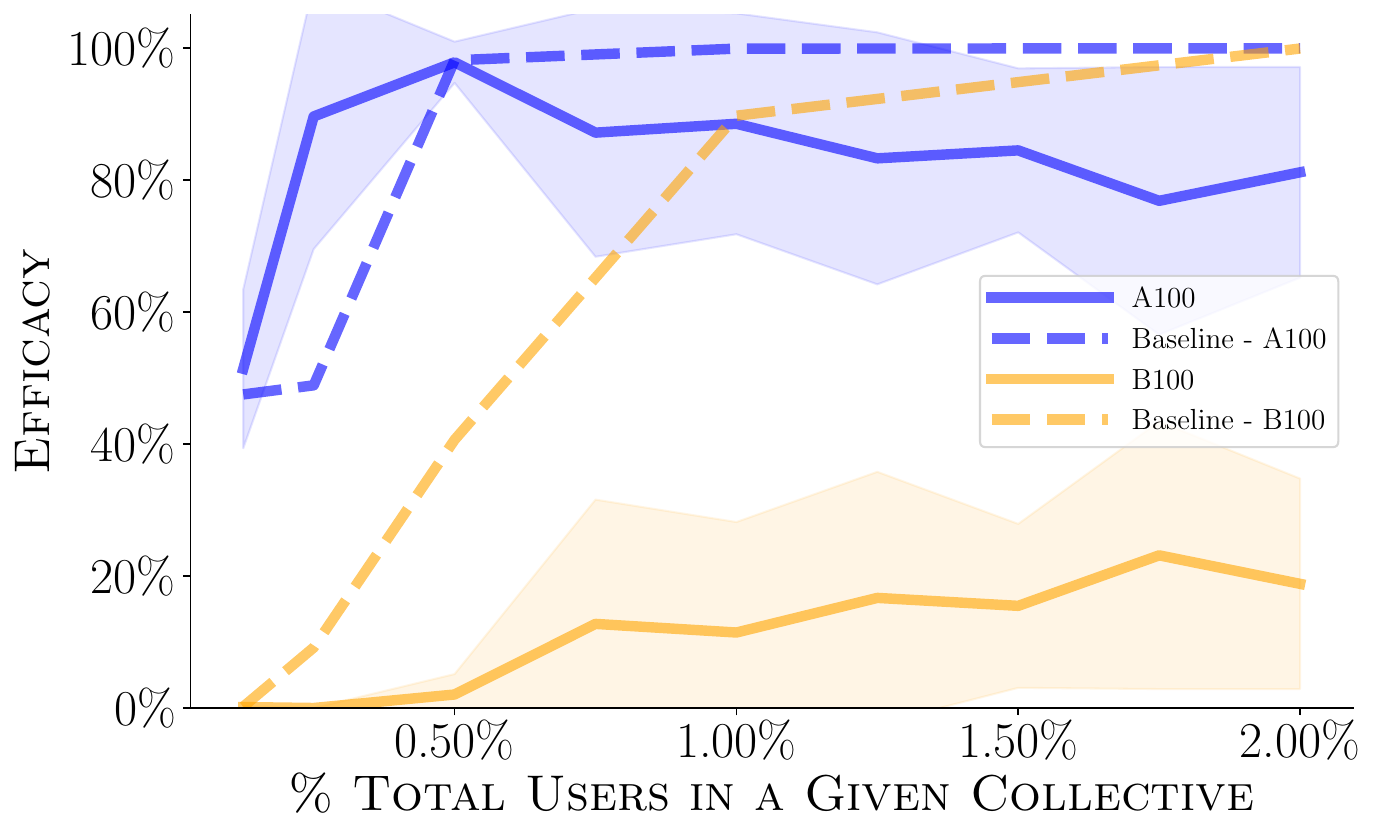}
    \caption{Different targets, Same chars}
        \label{fig:special_char_same_char}
    \end{subfigure}
    \begin{subfigure}{0.31\textwidth}
        \centering
        \includegraphics[width=\linewidth]{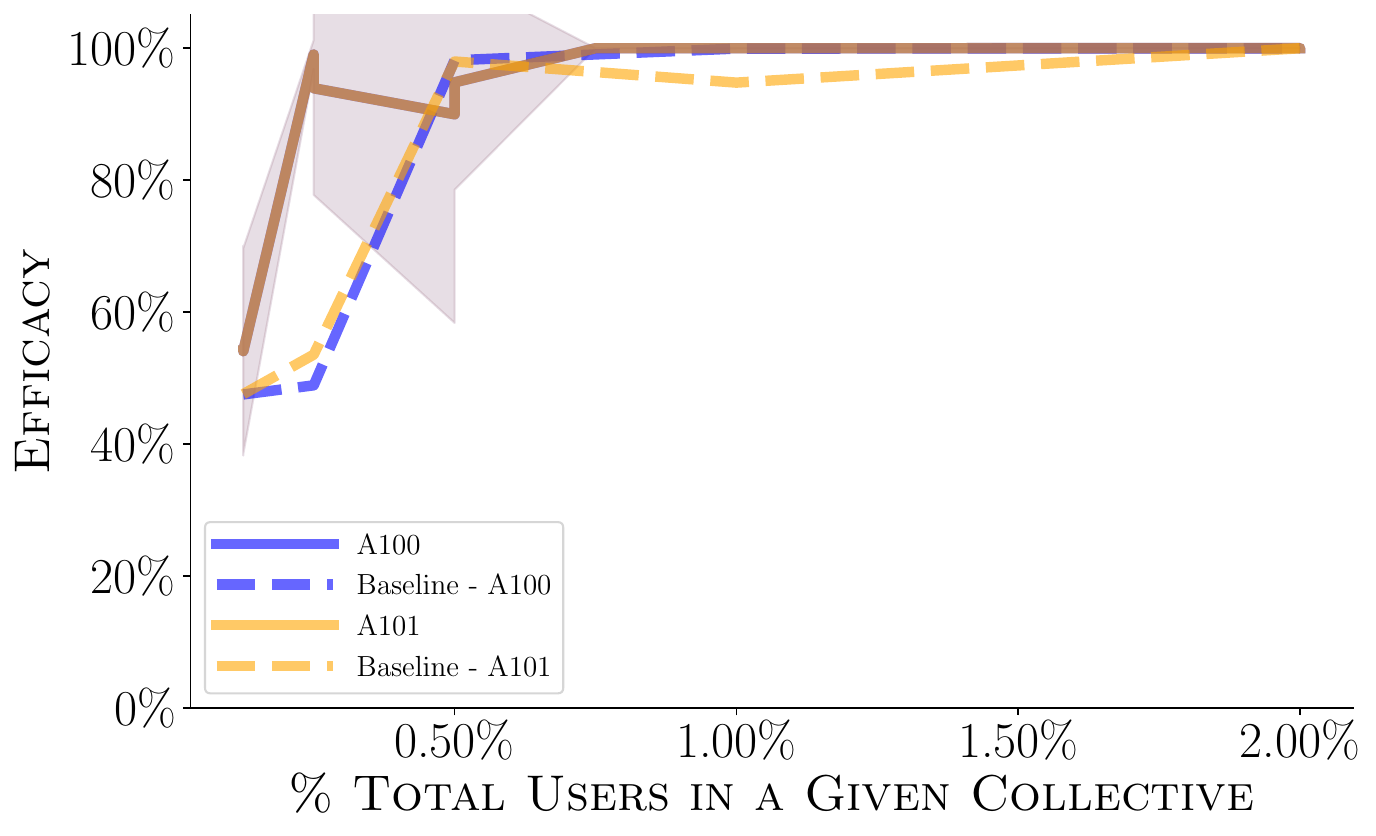}
        \caption{Same targets, Different chars}
        \label{fig:special_char_same_target}
    \end{subfigure}

    \caption{Multiple Collective Action in the Resume Modification Task. Two collectives, each with their own  strategy, insert a character to cause resumes with this character pattern to be classified to a given target class. Collectives are labeled [Letter][Number] where [Letter] corresponds to a target class and [Number] corresponds to a specific character used (mapped in \Cref{sec:appendix_classification}). Each row represents a different character set. The $x$-axis is the $\%$ of population participating in a given collective. The $y$-axis is the efficacy, in this case, the top-one accuracy on predicting the collective's target class. The dashed line represents the baseline efficacy of the collective acting alone on the system. The solid lines show the efficacy of one of the collectives when two of them are acting on the system. We can see behavior when the presence of another collective can be helpful (\ref{fig:normal_char_two_seperate} solid lines above dashed lines) where they can be antagonistic (\ref{fig:normal_char_same_char}, \ref{fig:special_char_two_seperate} \ref{fig:special_char_same_char} solid lines below dashed lines) or minor impact (\ref{fig:normal_char_same_char}, \ref{fig:special_char_same_char}). Most notably, while Figures \ref{fig:normal_char_two_seperate} and \ref{fig:special_char_two_seperate} are the same scenario, just with different characters, they produce very different outcomes (helpful interaction vs harmful) potentially due to the characters used in \ref{fig:normal_char_two_seperate} appear in non-modified data while the characters used in \ref{fig:special_char_two_seperate} only appear in the modified data.} 
    \label{fig:normal_char_langauge}
\end{figure*}

\Cref{fig:normal_char_langauge} shows the results with varying levels of participation (users chosen uniformly at random and the two collectives having equal sizes). The outcome is the efficacy: the probability that a user with this modification strategy is classified to the intended target. The $x$-axis represents the percentage of users participating in a given collective. The dashed lines represent the baseline when only one collective was acting on the model, while the solid represents the individual collective's efficacy in the presence of each other. 

In the first column, we see in \Cref{fig:normal_char_two_seperate} an instance of another collective acting on the target can be helpful (A0 vs B1 where the solid lines are above the dashed lines) especially in the low participating scenario. However, in the \Cref{fig:special_char_two_seperate} (A100 vs B101) the presence of another collective is harmful even when both collectives use different strategies and target different classes (solid lines below the dashed lines). One possible reason for this is that \Cref{fig:normal_char_two_seperate} the collectives uses more standard characters that exist (infrequently) in the unmodified dataset, while in \Cref{fig:special_char_two_seperate} each strategy uses never before seen characters. Indeed for this \texttt{distilbert-based-uncased} tokenizer, we find that, even though the characters themselves are distinct, the tokenizer treats the ``100" and ``101" characters as identical. This results in the model treating the two unique characters as a similar signal. This creates a conflict in this context.
The second column shows in direct confrontation (using the same strategy but targeting different classes) that collectives both lose some efficacy. In the case where they both target the same class but employ different strategies, there is not a noticeable loss in performance. 

\begin{figure*}[t]
    \centering
    \begin{subfigure}{0.85\textwidth}
        \centering
        \includegraphics[width=\linewidth]{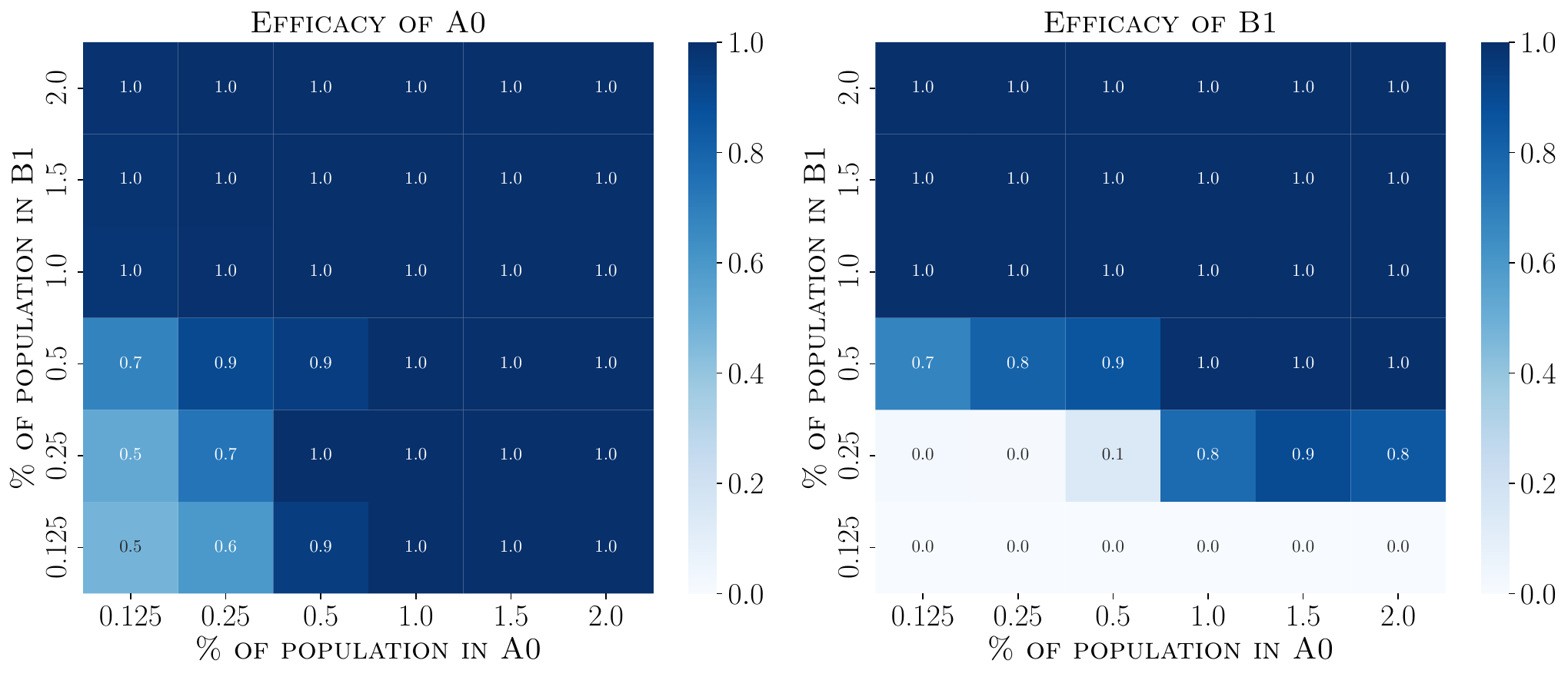}
        \caption{A0 vs B1 efficacy}
        \label{fig:varying_normal_char_two_seperate}
    \end{subfigure}
    \begin{subfigure}{0.85\textwidth}
        \centering
        \includegraphics[width = \linewidth]{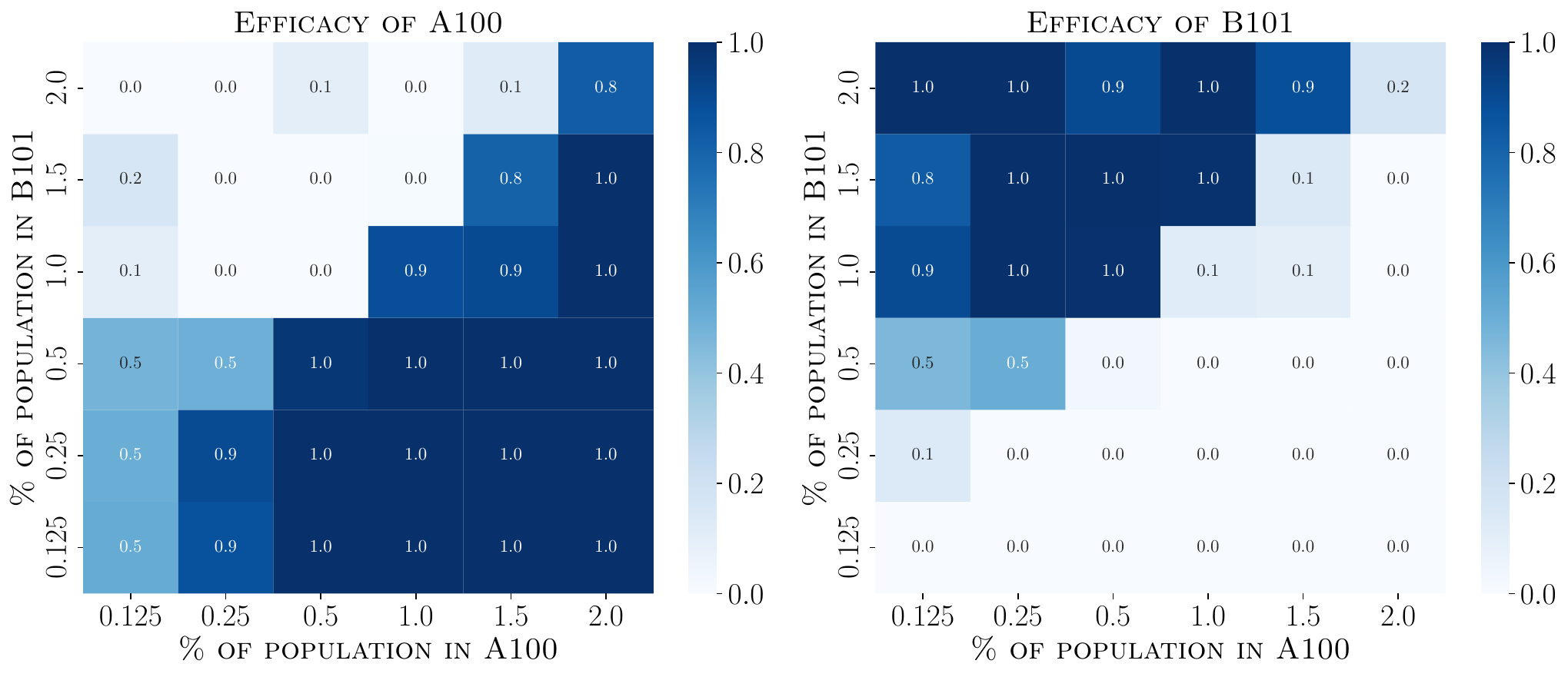}
    \caption{A100 vs B101 efficacy}
        \label{fig:varying_special_char_two_separate}
    \end{subfigure}

    \caption{Role of varying sizes. Here, there are two different strategy sets where, for each scenario, the strategy and the target used by each collective is distinct. The $x$-axis shows the percentage participation of the first collective, while the $y$-axis is the percentage participation in the second collective. Each square represents a single collective's efficacy (denoted in the title) (averaged across 5 trials). We see that, while there is some effect of collective sizes in the lower levels of participation in ~\ref{fig:varying_normal_char_two_seperate}, with sufficient amount of participation, both collectives get nearly full efficacy. This is in contrast to \ref{fig:varying_special_char_two_separate} where the two collectives act antagonistically. In particular, B101 requires nearly 2x participation compared to A100 to achieve its objective.} 
    \label{fig:varying_sizes}
\end{figure*}

\textbf{Role of Collective Size:}
We further examine how varying the sizes of each collective can affect efficacy. In~\Cref{fig:varying_sizes} we see that, while there is some effect of collective sizes in the lower levels of participation in~\ref{fig:varying_normal_char_two_seperate}, with enough participation, both collectives get nearly full efficacy. This is in contrast to \ref{fig:varying_special_char_two_separate} where the two collectives act antagonistically, notably we see explicit tradeoffs across varying levels of participation where one collective performs well while the other collective suffers. In particular, A100 outperforms B101 when at equal participation, collective B101 requires nearly 2x larger participation than A100 in order to be successful. 
This scenario is noteworthy as both the strategy and target these collectives operate on are distinct. As in the previous example, the distinctiveness of both the characters in this scenario may inadvertently cause conflict.

\subsection{RQ2: Impact of Heterogeneity}
\label{subsec:recsys_results}

\textbf{Experimental Details:}
We defined the high-level experimental setup in \Cref{sec:recsys_experiment}. For the specific instantiations for this experiment,  we set the number of collectives $C = 2$, which are selected from $Q=10$ number of clusters. We let the number of items picked for targeting $V = 10$ and evaluate the HR@K where $K=10$. 
We set $N = [10, 20, 50]$ to examine the impact on collective size. 
while having $p$ (propensity) vary from $[0.1, 0.25, 0.5, 0.75, 1]$. We use \textit{Surprise} ~\cite{Hug2020} as our implementation and perform grid search, across number epochs, learning rate and regularization with standard 5-fold cross validation to do parameter selection for each iteration of simulation. We use MovieLens 100k ~\cite{movielens2015} as the dataset where the users can either promote items (change rating to $5$) or demote items (change to rating $1$). We perform each of these experiments $100$ times. We use $L_2$ distance/k-means to determine clusters in one set of experiments and cosine distance/k-medoids in another set of experiments for robustness. To choose the two collectives' centroids, we consider two methodologies: one where we perform random selection of the seed clusters and one where we choose the two clusters that have the maximum distance from each other as the seed. For the results in the main body, we present the case where cosine distance paired with k-medoids are used for determining clusterings and the two collectives cluster's centroids are chosen to have maximum distances, we present the other scenarios in the ~\Cref{sec:appendix_Single}, \ref{sec:appendix_multiple} .
We run these experiments on a campus CPU cluster. $100$ iterations of any single parameter choice set take roughly four hours to run.

\textbf{Evaluation:}
We use the constructiveness score  $CT(c_i, c_j) = \frac{g_{i}(\tilde{\mathbf{X }}_i | \hat{\theta}_{i \wedge j})} {g_i(\tilde{\mathbf{X}}_i | \theta') } -  \frac{g_{i}(\tilde{\mathbf{X}}_i | \theta_{i})} {g_i(\tilde{\mathbf{X}}_i | \theta')}$
for both $CT(c_1, c_2)$ and $CT(c_2, c_1)$. In other words, we examine what happens to $c_1$'s objective in the presence of $c_2$ compared to when there's just $c_1$  operating in the system and vice versa. A ``promoter'' collective will want a positive score, while a ``demoter'' collective will want a negative score.

\begin{figure}
        \includegraphics[width=\linewidth]{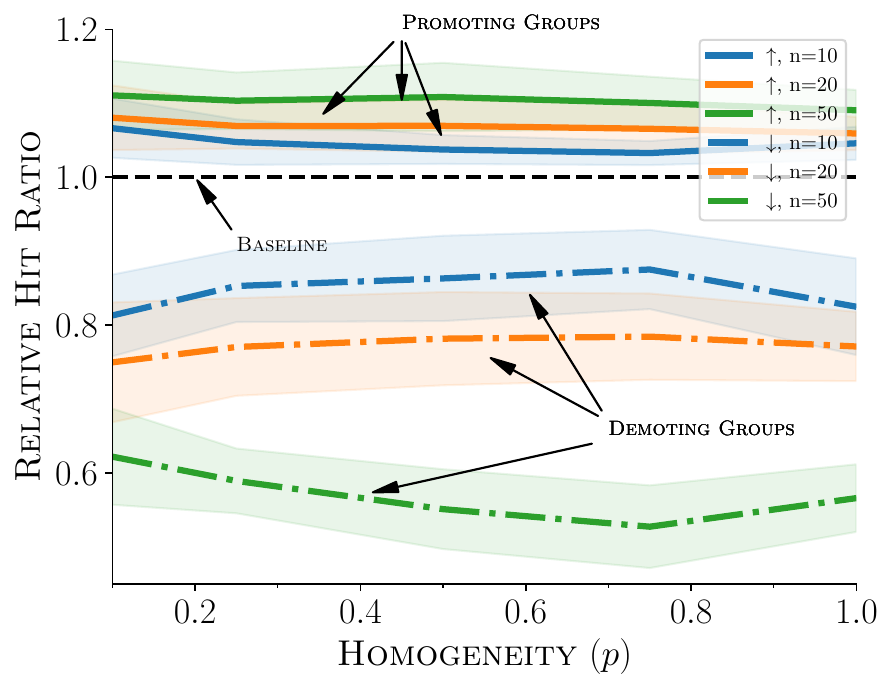}
        \caption{Impact on collective size and homogeneity in changing in HIT ratio for a single collective scenario. The means and the standard deviations of the relative HIT ratio are plotted. The $x$-axis represents homogeneity as measured by the sampling propensity ($p)$, a higher $p$ means that the members of the resulting collectives are more similar. The $y$-axis is the relative HIT ratio, in other words, how much higher/lower are the rankings of a collective's item when acting on the system vs no action. Blue represents $n=10$, orange is $n=20$ and green is $n=50$. The solid lines represent demoting collectives, while dashed lines represent promoting collectives. Collective size plays a much larger influence than homogeneity, especially for demoting groups; homogeneity plays a secondary influence.}
        \label{fig:single-cosine-max-main}
\end{figure}

\textbf{Single Collective Settings:}
As a preliminary study, we investigate how collective homogeneity and collective size impacts the ability to affect the hit ratio. We plot the relative hit ratio and show one of the design choices (grouping) and defer the rest for \Cref{sec:appendix_Single}. We find that the collective size has a first order impact on the ability to promote or demote rankings, while homogeneity has a second order impact. In \Cref{fig:single-cosine-max-main} we see that changing the collective size has a more pronounced impact than homogeneity. Regarding homogeneity ($p$), we see for demoter archetypes that the largest impact does not occur at $1$. Rather, it ranges between $0.1 - 0.75$; in other words, fully homogeneous collectives tend to do worse in effecting their objectives while collectives with varying levels of homogeneity are more effective in achieving their objective. The impact of homogeneity is more pronounced with demoter archetypes, the impact of homogeneity on the promoter archetype is minimal.

\begin{figure}
        \centering
        \includegraphics[width=\linewidth]{
        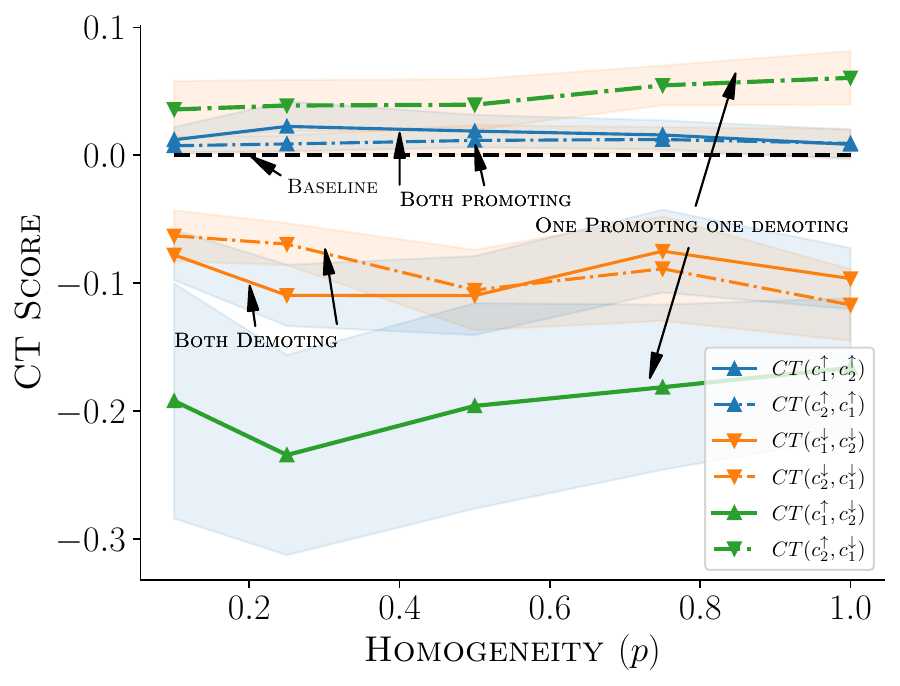}
    \caption{Constructive score vs. sampling propensity for a two collective scenario involving a recommender system. The constructiveness score between $c_1$ and $c_2$ measures the change in $c_1$'s objective with the presence of both $c_1$ and $c_2$ in the system compared to just $c_1$ in the system. Blue lines show a two ``promoting'' collectives scenario, while orange shows two `demoting'' collectives scenario. Green shows a scenario where there is promoter collective (solid line with $\blacktriangle$) and a second demoter collective (dashed lines with $\blacktriangledown$). Shaded regions are 1$\sigma$ ranges for the $100$ simulations. We see that the blue lines have a positive CT score, meaning that having both collectives act in the system is beneficial to their goal of boosting rankings. We see the orange lines are negative, meaning that having both collectives in the system is beneficial in their goal of demoting rankings.
    However, when one collective is trying to and the other is trying to decrease, we see that the collective that is trying to promote items (green $\blacktriangle$) sees a negative CT score and a collective that trying to demote items (green $\blacktriangledown$) has a positive CT score. These collectives are interfering with each other.}
    \label{fig:constructive}
\end{figure}

\textbf{Impact of multiple collectives:}
In \Cref{fig:constructive} we see what happens when both collectives attempt to promote different sets of items (blue), demote different sets of items (orange) and one collective promotes and another collective demotes (green).
The $x$-axis is the homogeneity defined by sampling propensity $p$, while the $y$-axis is the constructiveness score. We see that the blue lines have positive constructiveness score as desired. We also see that the orange lines (both collectives demoting) have negative scores as desired.
The $p$ that achieves better outcome generally doesn't occur at the extremes, suggesting that collectives that are more heterogeneous are more effective at making a change.

When multiple collectives are demoting ratings, it seems easier to achieve than when two collectives are increasing ratings. One possible reason is moving items away from the top 10 is easier than adding them into the top 10 by virtue that there are far more than 10 items to rank -- a small collective demoting ratings could be enough to bring it out of that list. When the collectives have different objectives, a conflict arises. In green, we see a scenario with a promoter archetype ($\blacktriangle$) and a demoter archetype ($\blacktriangledown$). In this case, the first collective would want their CT score to be positive, while the second collective would want it negative. However, it is the opposite. This implies that the two collectives are actively hindering each other. Given these two collectives have centroids that are maximally far apart, one may expect minor interference. However, this graph refutes this notion. This interference conveys the importance in understanding the effects of multiple collectives acting on algorithmic systems and how collectives can help or hinder each other.

\subsection{Summary of Results}
The results demonstrate that two collectives acting on systems can have complementary or antagonistic impact. We see the efficacy of algorithmic collective action increases as the complexity of models increase; a collective in the linear case had a relatively smooth, muted impact while in the language model setting, a minor change had a much more profound impact. A priori it may be difficult for collectives to anticipate which combinations of strategies can be complementary or antagonistic for complex models (\textit{e.g} models can interpret seemingly different behaviors as the same). Sizes of the collective are an important factor, which can allow one collective to nullify the impact of another collective's collective action. A collective's heterogeneity can play a second order influence on efficacy role for certain archetypes (demoters).

\section{Discussion}

\textbf{Role of Multiple Objectives:} Figure \ref{fig:constructive} demonstrates cases in which collectives assist each other (when collectives both promote) and interfere with each other (when one collective promotes and another demotes). While we cannot make universal claims that this pattern will always hold, our results above provide a concrete takeaway for both organizers of collective action and AI developers: interactions between collectives can have large impacts on collective action success and overall system performance. Furthermore, by using our experimental results or conducting similar experiments, both organizers and developers can reason about likely interactions.

\begin{figure*}[t]
    \centering
    \begin{subfigure}{0.45\textwidth}
        \centering
        \includegraphics[width=\linewidth]{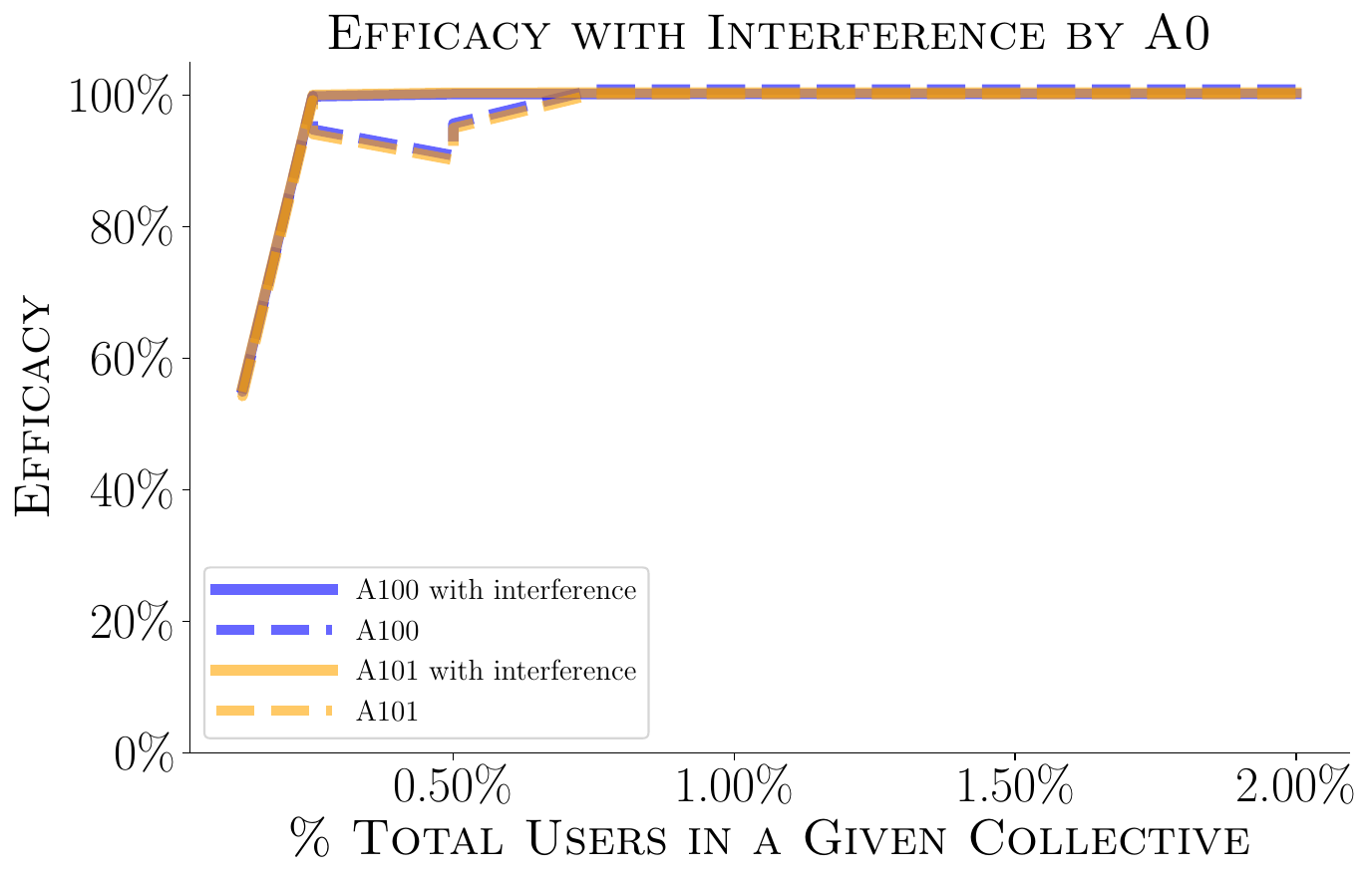}
        \caption{Effect of A0 with A100, A101 present} 
        \label{fig:interference_SSS}
    \end{subfigure}
    \begin{subfigure}{0.45\textwidth}
        \centering
        \includegraphics[width = \linewidth]{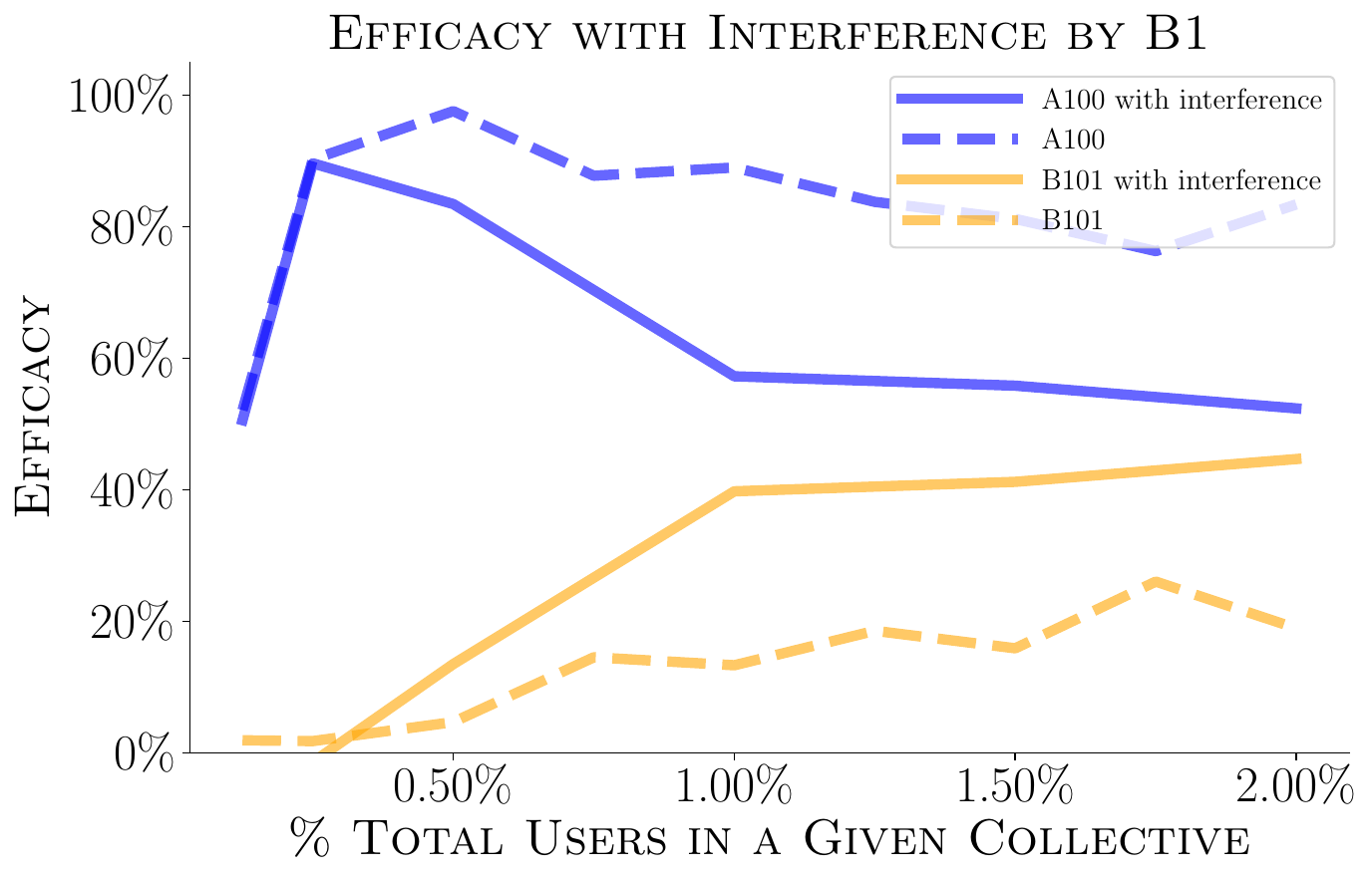}
    \caption{Effect of B1 with A100, B101 present}        
    \label{fig:interference_SWW}
    \end{subfigure}

    \caption{Role of a third collective on efficacy of the other two collectives. We see different types of potential impacts. The left shows minor impact with a third collective present. This may be because all the collectives are targeting the same class (A).  The right figure shows a decline in efficacy for the blue collective while an increase in efficacy for the orange collective. The interfering collective has the same target as the orange collective - the presence of the new collective boost B101 while weakening A100.} 
    \label{fig:three_collective}
\end{figure*}

\textbf{More than Two Collectives:}
Here we highlight some of the potential behaviors and challenges when analyzing more than two collectives.
To motivate future work and understand some of the complications that come with three collectives, here we consider the effect of a third collective participating in the classification example seen in \Cref{subsec:classifcation_experiment}. We consider the base case where two collectives are taking action. We then consider what happens to the first two collectives' efficacy when a third collective participates. In \Cref{fig:three_collective}, we plot the efficacy of two collectives in the presence of a third collective (solid lines). In \Cref{fig:interference_SSS} we see minor impacts, while in \Cref{fig:interference_SWW} we see a more pronounced impact. As the number of collectives increases, where each collective employs different strategies and targets, these interaction effects can become more difficult to understand or isolate. However, this remains an important area to characterize further. Earlier, we introduced the notion of a "constructiveness score", as a reminder, $CT(c_i, c_j)$ measures the impact that collective j has on collective i's objective. One way to extend this measure would be to compute pairwise constructiveness scores across two groups. Another method could be to employ a VCG style calculation, which would compute the harm to objective values of the other collectives when $j$ is present vs the objective values when $j$ is not present \cite{roughgarden2010algorithmic}. These measures can quantify a collective's impact against any number of others.

\textbf{Sanctions and Trustworthiness:}  In our experiments, all collective members act unilaterally (e.g., all set their ratings to 5-stars for a target set of items). However, the process by which collective members ensure that everyone acts in accord with the stated objective may involve disagreement and noise. For example, if a collective in a movie recommendation scenario wants to improve the ratings of romantic comedies, some users may interpret which movies count as romantic comedies differently. Classical collective action work ~\cite{ostrom1990governing} considers the ability of a collective to punish individuals in the collective when they do not behave to benefit the collective as a whole. Future experiments might engage with sociological work on collective action and social dilemmas \cite{heckathorn1996dynamics,kollock1998social}.

\textbf{What is Malicious Collective Action?} 
Our work contributes to an ongoing discussion around what kinds of collective action count as malicious \cite{etter2021activists, vincent_can_2021, vincent_data_2021}. The common framing in the adversarial ML literature frames those seeking to strategically act upon or against ML systems as ``attackers'' and those who prevent or fix this as ``defenders". This framing makes an implicit value judgement that ML system deployment is good for everyone and worth defending. However, these systems have biases, weaknesses, and goal misalignment from those who use this system. Some ``adversarial attacks" can help improve outcomes \cite{feng2022has}.
We do not claim that all collective action is net beneficial or that all ML systems are ``bad", but that collective action is core to understanding model behavior and could help mitigate ML harms. We recognize that some of these insights could be used for more nefarious means, underlying the need for more transparent algorithms and platforms to help identify nefarious use.

\textbf{Limitations and Future Work:} 
Our experiments looked at a subset of model types (BERT-based classification and classical matrix factorization). Exploring more complex models (\textit{e.g.} large language models), different and larger datasets/task, could show different implications on between-collective interaction. Prior work \cite{gupta2024fragilegiantsunderstandingsusceptibility} suggests that the ability to influence a model with small amounts of data scales with model complexity. Given our results on a smaller language model yielded stark results, we would expect that exporting this framework to a larger model would yield similarly interesting results. Further developing empirical and theoretical results with three or more collectives is another important line of investigation we leave for further work. 

As mentioned earlier, another avenue to expand these experiments would be to more fully simulate the social factors involved in collective formation and collective governance, potentially with network based formation. This may require richer data not typically available in open academic datasets or making various platform specific assumptions. This would involve further modeling individual motivations and action availability, accounting for how people respond to feedback from collective action, and modeling more than two or three collectives at once. 

Further nuanced behaviors between collectives could be explored, including multi-collective membership, goal sharing between collectives or collectives forming strategies based on other collectives. While these are promising avenues, this would further require deeper assumptions about how collectives engage organizationally with each other.

\section{Conclusion}
We introduced the first of a kind framework for analyzing data-driven collective action with multiple collectives. This framework invites exploration into settings in which multiple collectives, each  with their  objectives, attempt to influence a machine learning model for a specific outcome. These elements include differing objectives, strategies, collective size, and homogeneity. We demonstrated the utility of this framework by conducting experiments in both language classification and recommendation domain in a two collective scenario. In both contexts, we found evidence of interaction effects, potentially quite large, that can unintentionally help or hinder groups when performing algorithmic collective action. We see cases where a collective that was able to achieve near 100\% efficacy in their objective when acting alone could drop to nearly $25\%$ efficacy in the presence of another collective (\textbf{RQ1}). We also find that group heterogeneity may play a secondary impact in collective efficacy for demoter archetypes (\textbf{RQ2}). These interactions underscore the need to understand how this kind of strategic behavior affects systems and outcomes more broadly. Our work recognizes the role and impact that individuals working together can have on large algorithmic systems. It underscores how collectives can leverage their data to get better outcomes, even when algorithms are not transparent. It also highlights the importance of those providing data to models and how these individuals may work together to better promote their own interest in the face of large, algorithmic systems.

\begin{acks}
We would like to thank the many people who provided feedback on the paper: Leah Ajmani, Naina Balepur, Ti-Cheng Cheng, Brent Hecht,  Andy Lee, and Kaivalya Rawal. We would also like to thank the anonymous reviewers for providing feedback and suggestions.

\end{acks}

\bibliographystyle{ACM-Reference-Format}
\bibliography{ref}

\appendix
\newpage
\clearpage

\section{Linear Case} \label{sec:appendix_linear}
As discussed in the main body, we examine a simple logistic regression case using the Adult Income Dataset to predict high or low income. In this dataset, each person in the dataset has an occupation. For this experiment, we consider a two collective scenario where one collective wants a specific occupation A (Craft-repair) to be classified as high income, while another collective wants a different occupation B (Exec-managerial) to be classified as low income. 

\textbf{Collective Construction:} To construct collectives, we first partition the data into three clusters, $q_{A}$ representing all datapoints that have occupation A, $q_{B}$ representing all data points that have occupation B and $q_{R}$ which have the remaining data points. To see the effects of collective size, for collectives targeting $A$ we sample just from $q_{A}$ and for collectives targeting $B$ we sample just from $q_{B}$ to create $c_{A}$ and $c_{B}$. To achieve their objective, everyone in $c_{A}$ changes their label to positive and everyone in $c_{B}$ changes their label to negative. This modified data is then used in training the logistic regression classifier.

\textbf{Evaluation:} Success for $c_{A}$ is measured by whether test datapoints with class A are classified positively. For $c_{B}$ is it measured whether test datapoints with class B are classified negatively.

\textbf{Impact of Multiple Collectives as size:}
We examine for varying levels of participation in each group (noting that each of these groups represents around 12-13\% of the overall dataset).

\begin{figure}
        \includegraphics[width=0.5\textwidth]{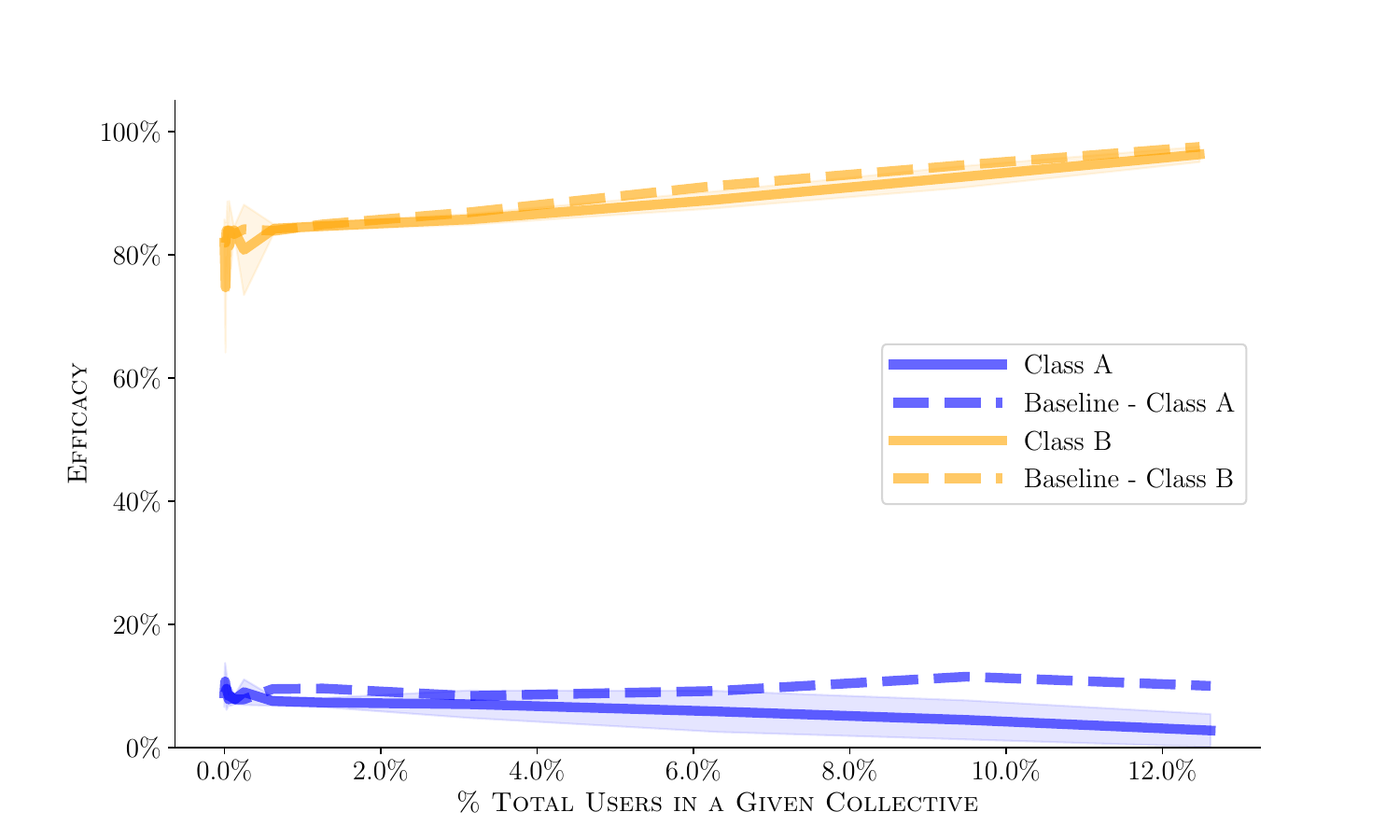}
        \caption{Collective Action in Linear Setting. Two groups target two separate classes (A and B) where the collective targeting class A wants to promote, while the collective targeting class B wants to demote. We plot the efficacy of achieving their objective, (for class A, a high score means the collective is successful in positive classification and a high score for class B means the collective is successful in getting negative classification). In this experiment, the first collective causes all data points corresponding to class A to have a positive label, while the second collective causes all data points corresponding to class B to have a negative label. Solid lines represent a scenario when both collectives are engaging, while dashed lines are when a single collective is engaging}
        \label{fig:equal_linear}
\end{figure}

\begin{figure*}
    \centering
    \includegraphics[width=\linewidth]{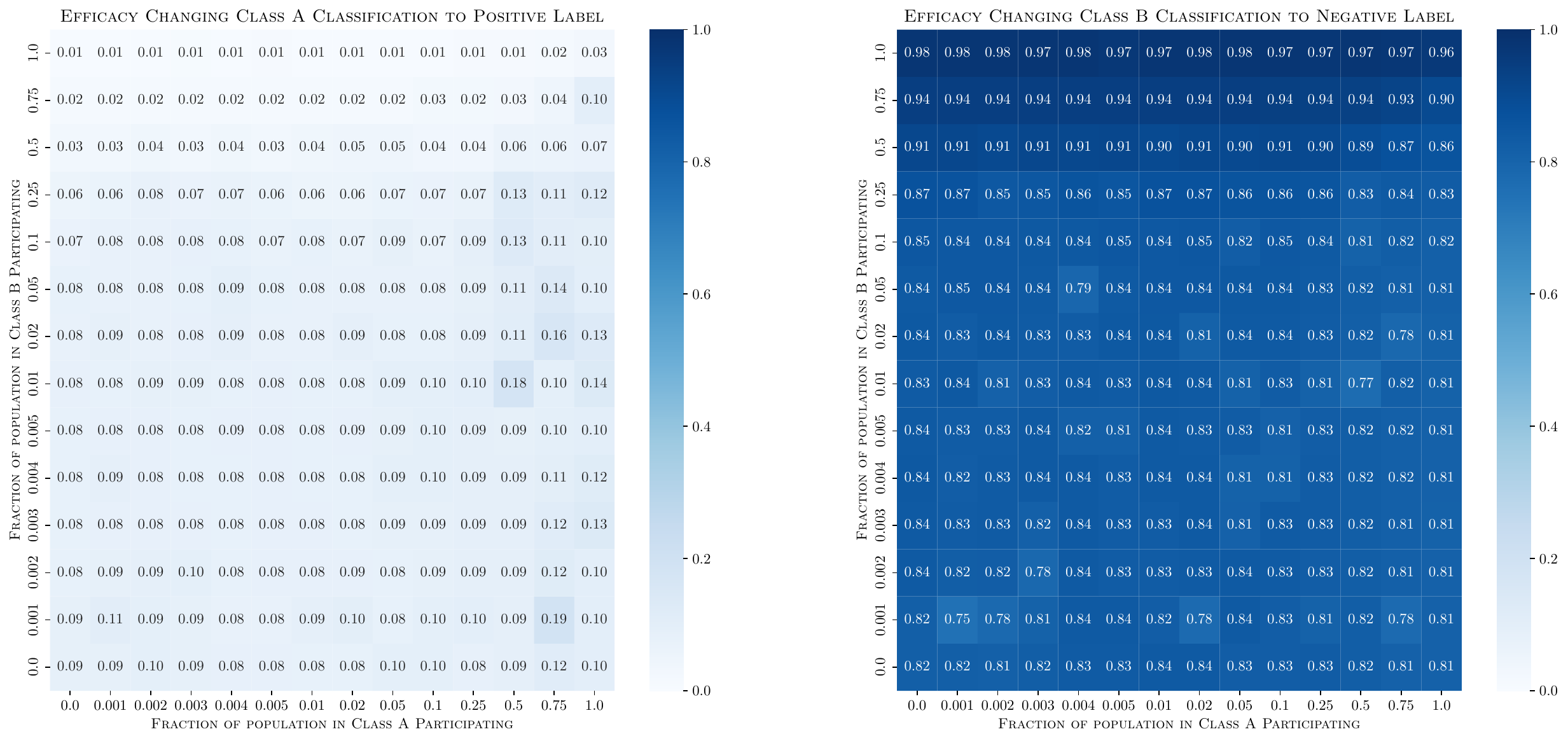}
    \caption{The role of varying level of participation on collective efficacy. Left Figure a group changing class A (Craft-repair) to positive, while the right is changing (Exec-managerial) to negative. Here we show the fraction of the given target class participating (hence a scale from 0 to 1). We show the efficacy in both cases. Compared to the non-linear cases, we see much less change and less direct interference across the board.} 
    \label{fig:linear-case}
\end{figure*}

\Cref{fig:equal_linear} has several takeaways. 1) In the baseline case (dashed lines), growing the collective size can increase the efficacy, but at a much more gradual pace (if any) compared to the systems shown in the main body. 2) The interaction effect (solid lines) is more muted than in the nonlinear cases. 
We see that for the group targeting A, even getting full group participation (everyone who has occuptation A is $\approx 12.5\%$ of the dataset) can at most reach 10-15\% efficacy, a relatively small change from 10\%. For class B, we see that the initial success rate is already at $80\%$ and still requires much of the group's participation to get to $>95\%$. Starting from a higher baseline can make this task slightly easier. We see that while there may be some interference in objectives, it is much more muted than in the non-linear cases and appears much more gradual. In \Cref{fig:linear-case} we further see the effect of varying the participation rate is less pronounced than in the non-linear case. 

\textbf{Impact of homogeneity: }
To test the impact of homogeneity, we first consider our clusterings $q_{A}, q_{B}$ and $q_{R}$. In the previous experiments, we assumed a fully homogeneous group $c_{A}$ only consisting of points from $q_{A}$ etc. To vary homogeneity, we consider sampling prosperity $p$. With probability, $p$ we sample for $c_{A}$ from $q_{A}$ and $1-p$ sample from $q_{R}$. We repeat the same for $c_{B}$.

To achieve their objective, each member of $c_{A}$ changes their occupation to be class A and sets their classification label to positive, while each member of $c_{B}$ changes their occupation to be of class B and sets their classification label to be negative. We perform the same training and consider the same evaluation criteria as described earlier.

\begin{figure}
    \centering
    \includegraphics[width=\linewidth]{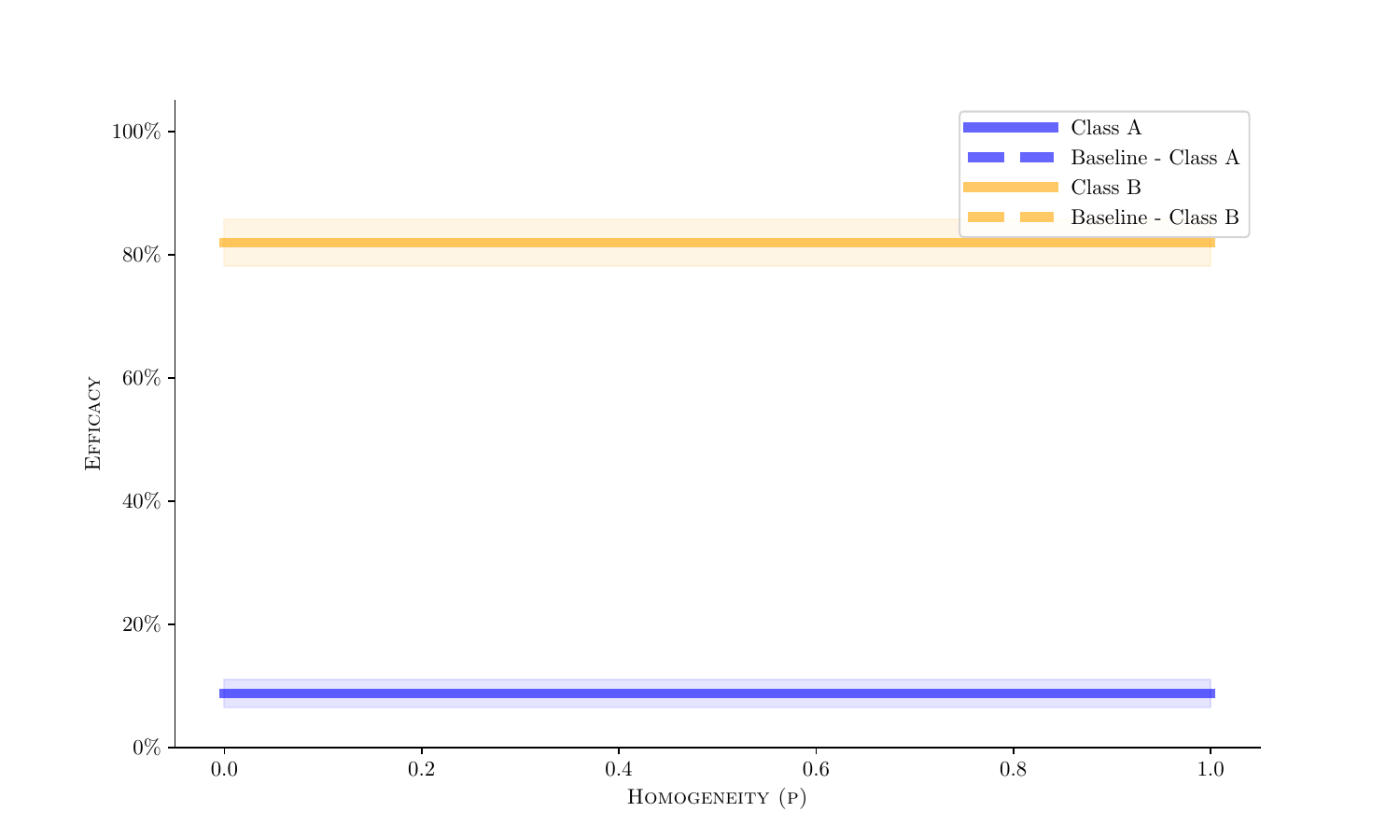}
    \caption{Effect of homogeneity on efficacy. Here, we fix the number of participants and vary homogeneity. In this context, a fully homogenous collective is one where all the data points modified were of target class while in a fully heterogeneous collective, all users are equally likely to be in the class. For users that were not originally a part of the target class, they change their career label to be the target class as well as change their target label to match the collective's goal. Here we see no impact on homogeneity, for either the single collective case (dashed lines) or the case when multiple collectives interact (solid lines)}
    \label{fig:linear-case-homo}
\end{figure}

\Cref{fig:linear-case-homo} suggests no impact due to homogeneity regardless of whether one group is participating or two groups are.

\section{Experimental Details for Language Classification}
\label{sec:appendix_classification}
Here we elaborate specific details about the resume classification experiments, specifically the strategy and target classes use. 

The strategy each group employs is inserting a specific character every 20 words as done in \cite{hardt_algorithmic_2023}.  In the main body, we labeled the target-character pairing as [Letter][Number]. We define the mapping in \Cref{tab:my_mapping}.

We use the resumes from \citet{jiechieu2021skills}. The data is given as a plaintext of resumes with the objective being multiclass classification of resumes into 10 jobs. We finetune a \texttt{distilbert\-based-uncased} \cite{Sanh2019DistilBERTAD}  model where we take $20,000$ resumes for training and $5,000$ for test. For each experimental trial, we create a collective based on the character they will insert, their target job classification, and the size of the group, expressed as a fraction of the training population. The combined modified training data, as well as the remaining clean data, are then used for finetuning. 
To evaluate the effectiveness of each collective, we take the $5,000$ test point and insert the unique character for the collective. The finetuned model then classifies the modified resume and we evaluated whether the predicted label matches the collective's intended target class.  We ran each experimental condition 10 times, where each experimental condition took ~30 minutes to complete. 
Experiments were ran on a ppc64le based cluster with V100 Nvidia GPUs.

\begin{table}[htbp]

    \centering
    \begin{tabular}{c|c}
         Group Prefix & Target Class \\
        \hline
         A & Software Developer \\ 
         B & Web Developer \\
         \vspace{.1in}
    \end{tabular}
    \centering
    \begin{tabular}{c|c}
         Group Suffix & Modifying Character \\
         \hline
         0 & \{ \\
         1 & \} \\
         100 & UTF-2E17 \\
         101 & UTF-2E18 \\
    \end{tabular}

    \caption{Information about collective definitions. For example, collective ``A0'' would represent a group targeting Software Developer using the \{ 
    for character modification in the resume task.}
    \label{tab:my_mapping}

\end{table}

\section{Alternative Metrics of Single Group Collectives in Recommender Systems}\label{sec:appendix_Single}

\begin{figure*}[t]
    \centering
    \begin{subfigure}[b]{0.47\textwidth}
        \centering
        \includegraphics[width=\textwidth]{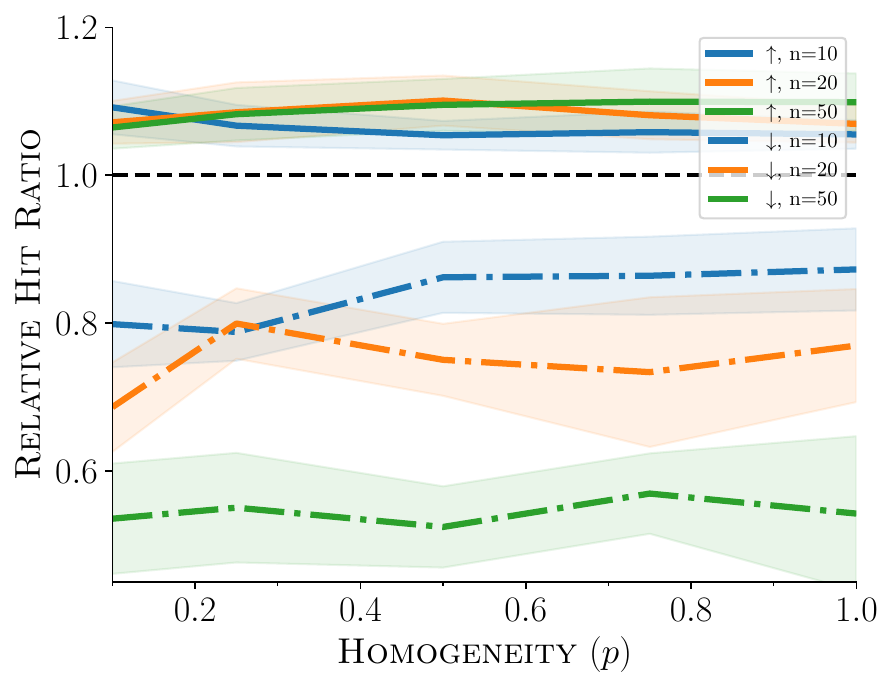}
        \caption{$L_2$/ K-means - Uniform}
        \label{fig:single-l2-random}
    \end{subfigure}
    \hfill
    \begin{subfigure}[b]{0.47\textwidth}
        \centering
        \includegraphics[width=\textwidth]{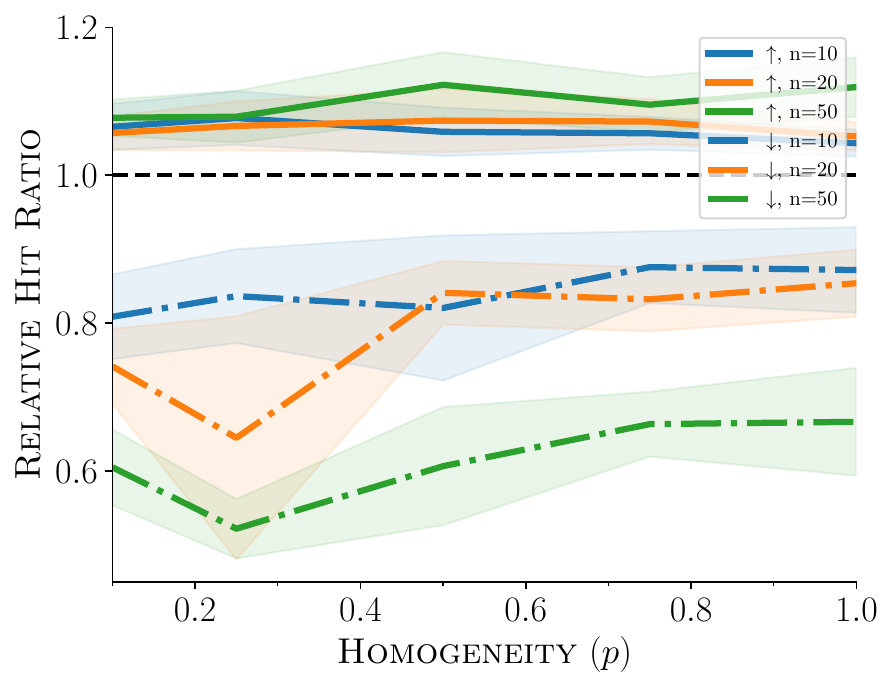}
        \caption{$L_2$/ K-means - Max distance}
        \label{fig:single-l2-max}
    \end{subfigure}
    \vskip\baselineskip
    \begin{subfigure}[b]{0.47\textwidth}
        \centering
        \includegraphics[width=\textwidth]{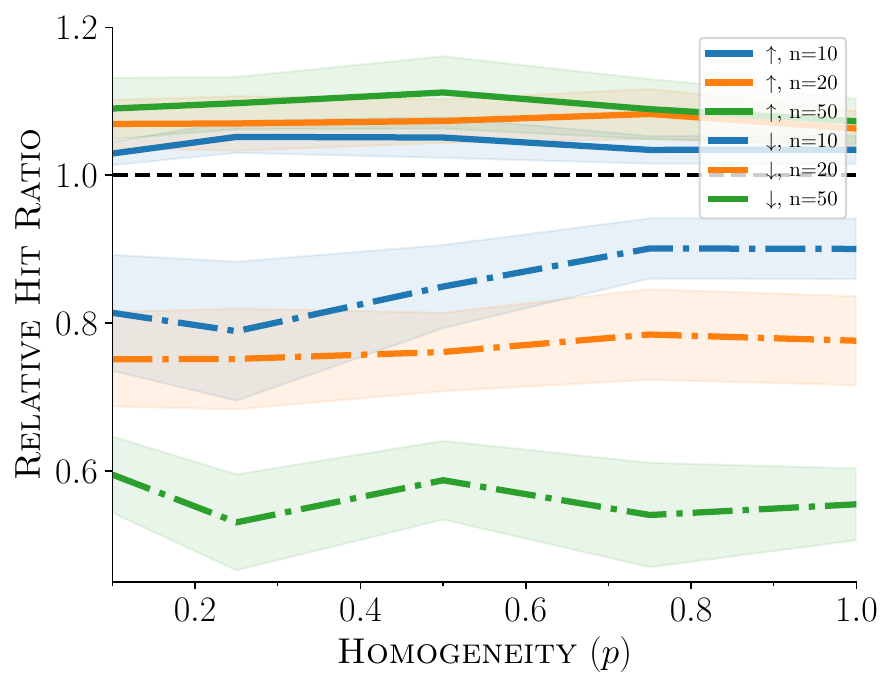}
        \caption{Cosine/K-Medoids - Uniform}
        \label{fig:single-cosine-rand}
    \end{subfigure}
    \hfill
    \begin{subfigure}[b]{0.47\textwidth}
        \centering
        \includegraphics[width=\textwidth]{figures/Group_size_impact_Metric=cosine-Grouping=kmedoids-Choosing=max_distance-both_directions_annotation-True.pdf}
        \caption{Cosine/K-Medoids - Max Distance}
        \label{fig:single-cosine-max}
    \end{subfigure}
    \caption{Impact on group size and similarity on changing  the HIT ratio for different choices of grouping/metrics (bottom right having appeared in the main body). The means and the standard deviations of the relative HIT ratio are plotted. The $x$-axis represents the sampling propensity, a higher sampling propensity means that the members of the resulting group is more similar. The $y$-axis is the relative HIT ratio, in other words how much higher/lower are the rankings of a group's item when acting on the system vs no action. Blue represents $n=10$, orange is $n=20$ and green is $n=50$. The solid lines represent demoting groups while solid lines represent promoting groups. Each figure represent different choices in collective construction -- the metric involved and the way the centroids are chosen. }
    \label{fig:single}
\end{figure*}

As discussed in the main body, we looked at the direct impact on size and group homogeneity on the relative hit ratio. The main body presented a single grouping/metric choice. Here we present the remaining. In ~\Cref{fig:single} we see that the size of the collective has a first order impact while homogeneity has a second order impact across metric types. Size is the primary factor for driving change in relative HIT ratio, we see that when $n=50$, it is the highest performing increasing variant ($>1$) or the best performing demoting variant $(<1)$

\section{Alternative Metrics for Multiple Group Collectives for Recommenders}\label{sec:appendix_multiple}

As discussed in the main body, we showed a single metric regime of multiple collective action (using cosine/K-Medoids to compute clusters and using the max distance criteria to determine centroids for the collectives). 
Here we present the figures for the remaining metric choices.

\begin{figure*}[t]
    \centering
    \begin{subfigure}[b]{0.45\textwidth}
        \centering
        \includegraphics[width=\textwidth]{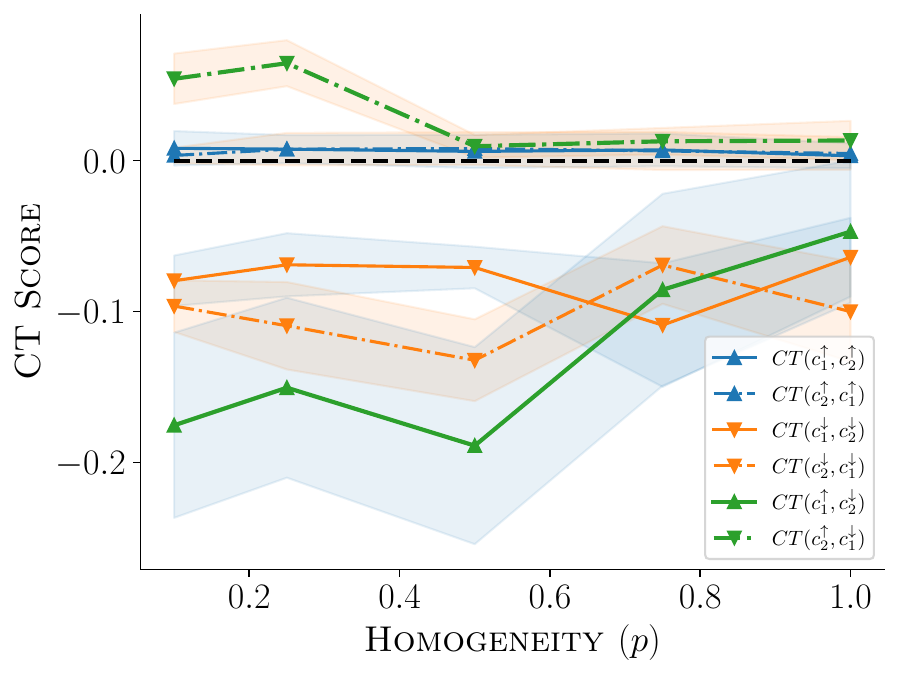}
        \caption{$L_2$ + K-means - Uniform}
        \label{fig:multiple-l2-random}
    \end{subfigure}
    \hfill
    \begin{subfigure}[b]{0.45\textwidth}
        \centering
        \includegraphics[width=\textwidth]{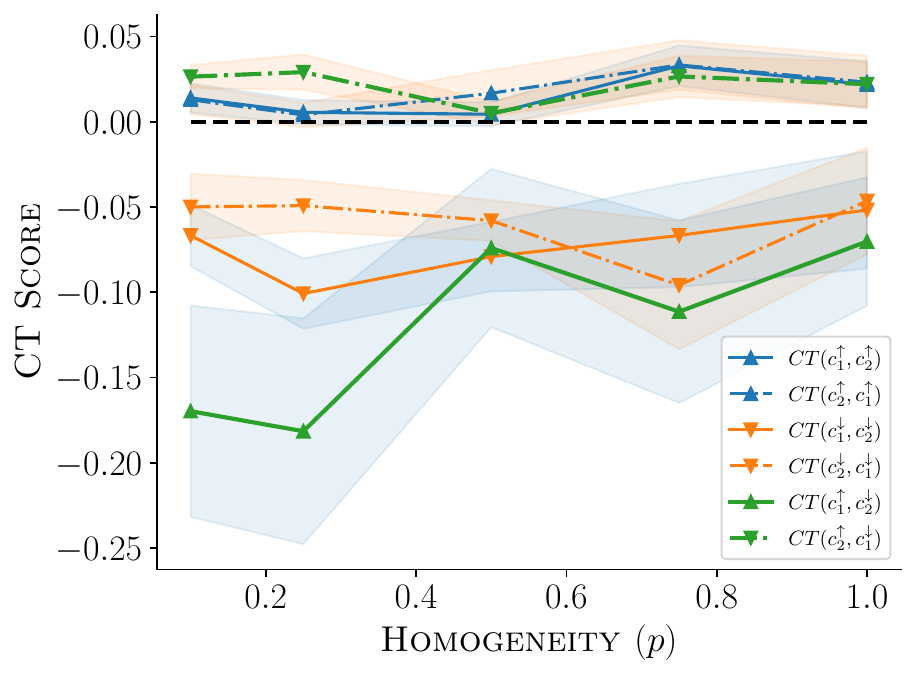}
        \caption{$L_2$ + K-means - Max distance}
        \label{fig:multiple-l2-max}
    \end{subfigure}
    \vskip\baselineskip
    \begin{subfigure}[b]{0.45\textwidth}
        \centering
        \includegraphics[width=\textwidth]{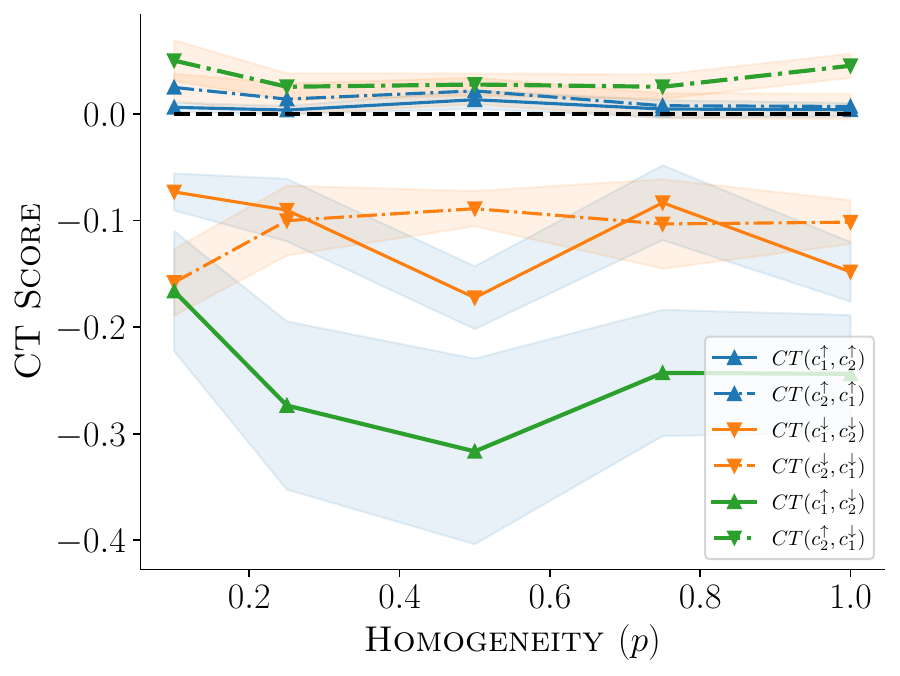}
        \caption{Cosine + K-Medoids - Uniform}
        \label{fig:multiple-cosine-rand}
    \end{subfigure}
    \hfill
    \begin{subfigure}[b]{0.47\textwidth}
        \centering
        \includegraphics[width=\textwidth]{figures/Constructivess_Group_size=50-Metric=cosine-Grouping=kmedoids-Choosing=max_distance-direct=all_annotate=True.pdf}
        \caption{Cosine/K-Medoids - Max Distance}
        \label{fig:multiple-cosine-max}
    \end{subfigure}
    \caption{Impact of multiple collective data action on recommender systems with different metric choices (bottom right having appeared in the main body). Homogeneity is on the x-axis while constructivess score is shows on the $y$-axis. Blue lines show a two ``promoting'' collectives scenario, while orange shows two `demoting'' collectives scenario. Green shows a scenario where there is promoter collective (solid line with $\blacktriangle$) and a second demoter collective (dashed lines with $\blacktriangledown$).}
    \label{fig:multi-metrics}
\end{figure*}

In \Cref{fig:multi-metrics}, we find a similar result as the main body,  that size tends to play a lager factor than homogeneity. In \Cref{fig:multiple-l2-max}, however, for the demoting $n=10$ and $n=20$ appear to tradeoff efficacy with differing $p$ though error bands are much larger.

\end{document}